# *I. The Isotopic Foldy-Wouthuysen Representation and Chiral Symmetry*

*V.P. Neznamov*[*]

*RFNC-VNIIEF, Sarov, Russia*

## Abstract

The paper introduces the isotopic Foldy-Wouthuysen representation. This representation was used to derive equations for massive interacting fermion fields.

When the interaction Hamiltonian commutes with the matrix $\gamma^5$, these equations possess chiral invariance irrespective of whether fermions have mass or are massless.

The isotopic Foldy-Wouthuysen representation preserves the vector and axial currents irrespective of the fermion mass value. In the Dirac representation, the axial current is preserved only for massless fermions.

In the isotopic Foldy-Wouthuysen representation, the ground state of fermions (vacuum) turns out to be degenerate, and therefore there is the possibility of spontaneously breaking parity (P - symmetry).

This study considers the example of constructing a chirally symmetric quantum electrodynamics framework in the isotopic Foldy-Wouthuysen representation. A number of physical processes are calculated in the lowest orders of the perturbation theory. Final results of the calculations agree with the results of the standard quantum electrodynamics.

[*] e-mail: neznamov@vniief.ru

# 1. INTRODUCTION

The mass term in the Dirac equation is known to mix right and left Dirac bispinor components and break chiral symmetry. Chiral symmetry is preserved only for massless fermions.

This study shows that for interaction Hamiltonians of the fermions with boson fields commuting with the matrix $\gamma^5$ one can construct a chirally symmetric integro-differential equation and a second-order equation equivalent to the Dirac equation. These equations can be written individually for left and right fermions, whether they possess mass or not.

The linear form of chirally symmetric equations for interacting fermion fields with respect to the operator $p_0 = i\frac{\partial}{\partial t}$ can be obtained using the Foldy-Wouthuysen transformation [1] in a specially introduced isotopic space, in which the Dirac equation is preliminarily wrote. In this case, chiral symmetry is also preserved irrespective of the fermion mass value.

In the Dirac representation, the vector current is preserved, whereas the axial current is preserved only for massless fermions.

In the introduced isotopic Foldy-Wouthuysen representation, both vector and axial currents are preserved irrespective of whether fermions possess mass or not.

From the three equivalent Dirac equations written in different ways in the isotopic space, one can obtain three chirally symmetric Foldy-Wouthuysen equations by the same isotopic Foldy-Wouthuysen transformation. These equations differ from each other in their physical content due to the features of the Foldy-Wouthuysen representation, which restricts the space of possible states of a Dirac particle. One equation describes both right and left fermions, and right and left antifermions. Two other equations describe either right fermions and left antifermions, or left fermions and right antifermions.

It follows from this that the ground (vacuum) state of fermions turns out to be degenerate in the isotopic Foldy-Wouthuysen representation. Along with the chirally symmetric “sea” of negative-energy fermions, there are chirally symmetric vacuums with disturbed P - symmetry, namely there is a vacuum with a “sea” of right fermions and a vacuum with a "sea” of left fermions. For fermions, this creates conditions for the spontaneous breaking of P - symmetry.

The above issues are discussed in Sections 2 – 5 of this paper. Basic features of the Foldy-Wouthuysen representation are considered in Section 1. Construction of a chirally symmetric quantum electrodynamics framework in the isotopic Foldy-Wouthuysen representation is considered as an example in Section 6.

The algorithm for deriving the fermion Hamiltonian with interaction in the isotopic Foldy-Wouthuysen representation is presented in Appendix 1.

Calculations of some quantum electrodynamics processes in the isotopic Foldy-Wouthuysen representation are presented in Appendix 2. Final results of the calculations agree with corresponding results of the standard quantum electrodynamics.

## 2. Basic features of the Foldy-Wouthuysen representation

As we known, the Foldy-Wouthuysen (FW) transformation is performed by means of the unitary operator $U_{FW}$ [1].

The wave function (Dirac field operator) and the Hamiltonian of the Dirac equation in this case transform in the following way:

$$p_0\psi = H_D\psi;$$

$$\psi_{FW} = U_{FW}\psi;$$

$$H_{FW} = U_{FW}H_D U_{FW}^{\dagger} - iU_{FW}\frac{\partial U_{FW}^{\dagger}}{\partial t}. \qquad (1)$$

$$p_0\psi_{FW} = H_{FW}\psi_{FW}$$

In expressions (1) and below, the system of units is $\hbar = c = 1$; metrics of the Minkowski space is taken in the form of $g^{\mu\nu} = diag\left[1,-1,-1,-1\right]$; $p^{\mu} = i\left(\frac{\partial}{\partial x_{\mu}}\right)$; $\psi(x)$, $\psi_{FW}(x)$ are four-component wave functions (field operators) in the Dirac and Foldy-Wouthuysen representations.

In the paper, Dirac matrices in the Hamiltonians $H_D$, $H_{FW}$ are used both in the Dirac-Pauli and in the chiral representation, which is widely used in the Standard Model.

In the Dirac-Pauli representation:

$\beta = \gamma_0 = \rho_3$; $\alpha^i = \beta\gamma^i = \rho_1\sigma^i$; $\gamma^i = i\rho_2\sigma^i$; $\gamma^5 = i\gamma^0\gamma^1\gamma^2\gamma^3 = \rho_1$; $\Sigma^i = E_{4\times4}\sigma^i$; $E_{N\times N}$ is the identity matrix $N\times N$; $\sigma^i$ are the 2x2 Pauli matrices; $\rho_1 = \begin{pmatrix} 0 & E_{2\times2} \\ E_{2\times2} & 0 \end{pmatrix}$; $\rho_2 = \begin{pmatrix} 0 & -iE_{2\times2} \\ iE_{2\times2} & 0 \end{pmatrix}$; $\rho_3 = \begin{pmatrix} E_{2\times2} & 0 \\ 0 & -E_{2\times2} \end{pmatrix}$.

In the chiral representation:

$\beta = \gamma_0 = \rho_1$; $\alpha^i = \beta\gamma^i = \rho_3\sigma^i$; $\gamma^i = \beta\alpha^i = -i\rho_2\sigma^i$; $\gamma^5 = i\gamma^0\gamma^1\gamma^2\gamma^3 = \rho_3$; $\Sigma^i = E_{4\times4}\sigma^i$.

For free motion with the matrices $\alpha^i$, $\beta$ in the Dirac-Pauli representation,

$$(H_0)_D = \boldsymbol{\alpha}\mathbf{p} + \beta m;$$

$$(H_0)_{FW} = (U_0)_{FW}(H_0)_D(U_0)^{\dagger}_{FW} = \beta E;$$

$$(U_0)_{FW} = R\left(1+L\right) = \sqrt{\frac{E+m}{2E}}\left(1+\frac{\beta\boldsymbol{\alpha}\mathbf{p}}{E+m}\right); \tag{2}$$

$$E = \sqrt{m^2+\mathbf{p}^2}, \quad R = \sqrt{\frac{E+m}{2E}}, \quad L = \frac{\beta\boldsymbol{\alpha}\mathbf{p}}{E+m}.$$

After the FW transformation, the Dirac equation

$$p_0\psi(x) = \left(\boldsymbol{\alpha}\mathbf{p} + \beta m\right)\psi(x) \tag{3}$$

transforms into the Foldy-Wouthuysen equation

$$p_0\psi_{FW}(x) = (H_0)_{FW}\psi_{FW}(x) = \beta E\psi_{FW}(x) \tag{4}$$

In the Foldy-Wouthuysen equation one can see clear asymmetry in space and time coordinates, although it is Lorentz-invariant in itself.

For the chiral representation of the Dirac matrices, expressions for the transformation operator $U_{FW}^{chir}$ and Hamiltonian $H_{FW}^{chir}$ are deduced from the expressions for $U_{FW}$, $H_{FW}$ with the matrices $\alpha^i$, $\beta$, $\gamma^5$ in the Dirac-Pauli representation with replacements of $m \leftrightarrow \boldsymbol{\Sigma}\mathbf{p}$, $\beta \leftrightarrow \gamma^5$ [2].
In particular, the Foldy-Wouthuysen equation for free motion in the chiral representation takes the form

$$p_0\psi_{FW}(x) = (H_0^{chir})_{FW}\psi_{FW}(x) = \gamma^5 E\psi_{FW}(x) \tag{5}$$

Solutions of free equations (4), (5) are plane waves of both positive and negative energy

$$\begin{aligned} \psi_{FW}^{(+)}(x,s) &= \frac{1}{(2\pi)^{3/2}}U_s e^{-ipx}; \\ \psi_{FW}^{(-)}(x,s) &= \frac{1}{(2\pi)^{3/2}}V_s e^{ipx}; \end{aligned} \tag{6}$$

$$U_s = \begin{pmatrix} \chi_s \\ 0 \end{pmatrix}; \quad V_s = \begin{pmatrix} 0 \\ \chi_s \end{pmatrix}; \; p_0 \equiv E.$$

In expressions (6), $x, p$ are four-vectors, and $\chi_s$ are normalized two-component Pauli spin functions.

In the presence of interaction with a common boson field, there is no closed Foldy-Wouthuysen transformation.

For stationary external fields, the general form of the exact FW transformation has been found by Eriksen [3].

Another direct way of deriving the Foldy-Wouthuysen transformation for the case of interaction of fermions with arbitrary (including time-dependent) boson fields has been developed in studies by the author (see [4] and overview paper [5]).

The transformation matrix and the relativistic Hamiltonian have been obtained in the form of a series in powers of the coupling constant

$$U_{FW} = \left(U_0\right)_{FW}\left(1+\delta_1+\delta_2+\delta_3+....\right)$$
$$H_{FW} = \beta E + K_1 + K_2 + K_3 + .... \tag{7}$$

In equalities (7) $\left(U_0\right)_{FW}$ is the *FW* transformation matrix for free Dirac particles; $E=\sqrt{\mathbf{p}^2+m^2}$ ; $\delta_1, K_1 \sim q;\ \delta_2, K_2 \sim q^2;\ \delta_3, K_3 \sim q^3;\ q$ is the coupling constant.

Besides the direct methods of transition to the Foldy-Wouthuysen representation, there are many step-by-step techniques for deriving FW equations and their Hamiltonians.

In particular, one of such techniques was used in the classical Foldy-Wouthuysen study [1] to derive a Hamiltonian for a Dirac particle in the presence of an external electromagnetic field in the form of a series in powers $1/m$.

The step-by-step techniques are generally valid for the first two iterations and are useful only for weak boson fields and their space-time derivatives [6].

In Refs. [4], [5], [7], the Foldy-Wouthuysen transformation was used for constructing the Standard Model and, in particular, for quantum electrodynamics. In order to account of interaction of real electron-positron pairs, the FW representation was modified, and specific processes of the quantum field theory were calculated.

Let us formulate the basic features of the Foldy-Wouthuysen transformation and representation:

1. In the Foldy-Wouthuysen representation, the Hamiltonian $H_{FW}$ is block-diagonal with respect to the upper and lower components of the wave function (field operator).

2. An obligatory condition for the transition to the *FW* representation for free motion and motion of fermions in static external boson fields is that either upper or lower components of $\psi_{FW}$ should be equal to zero (condition of Dirac wave function reduction) [6].

3. Because of the form of the basis wave functions (field operators) in the *FW* representation with zero upper or lower components the Foldy-Wouthuysen representation constricts the space of possible states of a Dirac particle. Special measures (modification of the *FW* representation) are required for getting back to the Dirac space of states.

## 3. Isotopic Foldy-Wouthuysen representation and chirally symmetric equations of motion of massive fermion fields

Let us consider the Hamiltonian density of a Dirac particle with mass $m$, which interacts with an arbitrary vector boson field $B^{\mu}(x)$

$$\mathcal{H}_D = \psi^{\dagger}\left(\boldsymbol{\alpha}\mathbf{p}+\beta m+q\alpha_{\mu}B^{\mu}\right)\psi=\psi^{\dagger}\left(P_L+P_R\right)\left(\boldsymbol{\alpha}\mathbf{p}+\beta m+q\alpha_{\mu}B^{\mu}\right)\left(P_L+P_R\right)\psi=$$

$$=\psi_L^{\dagger}\left(\boldsymbol{\alpha}\mathbf{p}+q\alpha_{\mu}B^{\mu}\right)\psi_L+\psi_R^{\dagger}\left(\boldsymbol{\alpha}\mathbf{p}+q\alpha_{\mu}B^{\mu}\right)\psi_R+\psi_L^{\dagger}\beta m\psi_R+\psi_R^{\dagger}\beta m\psi_L \tag{8}$$

In equalities (8) $q-$ is the coupling constant; $\alpha^i,\beta$ - матрицы Дирака, $\alpha^{\mu}=\begin{cases}1 & \mu=0\\ \alpha^i & \mu=i=1,2,3\end{cases}$; $P_L=\frac{1-\gamma_5}{2}, P_R=\frac{1+\gamma_5}{2}-$ are the left and right projection operators; $\psi_L=P_L\psi,\quad \psi_R=P_R\psi-$are the left and right components of the Dirac field operator $\psi$.

The Abelian case for the field $B^{\mu}(x)$ is considered for simplicity. If we consider the general case of a Dirac particle interacting with a non-Abelian boson field, the conclusions and results obtained in this study will be the same.

The Hamiltonian density $\mathcal{H}_D$ allows obtaining the motion equations for $\psi_L$ and $\psi_R$

$$\begin{aligned}p_0\psi_L&=\left(\boldsymbol{\alpha}\mathbf{p}+q\alpha_{\mu}B^{\mu}\right)\psi_L+\beta m\psi_R\\ p_0\psi_R&=\left(\boldsymbol{\alpha}\mathbf{p}+q\alpha_{\mu}B^{\mu}\right)\psi_R+\beta m\psi_L\end{aligned} \tag{9}$$

One can see that both Hamiltonian density $\mathcal{H}_D$ (8) and Eqs. (9) have such a form that the presence of mass in fermions results in mixing the right and left components of the field operator $\psi$, and chiral symmetry, and SU(2)xU(1) – symmetry is therefore preserved only for massless fermions.

### 3.1 Chirally symmetric equations for massive fermions in the Dirac representation

Now, let us question ourselves, whether it is possible to write chirally symmetric equations and their Hamiltonians for massive fermions.

It follows from Eqs. (9) that

$$\begin{aligned}\psi_L&=\left(p_0-\boldsymbol{\alpha}\mathbf{p}-q\alpha_{\mu}B^{\mu}\right)^{-1}\beta m\psi_R\\ \psi_R&=\left(p_0-\boldsymbol{\alpha}\mathbf{p}-q\alpha_{\mu}B^{\mu}\right)^{-1}\beta m\psi_L\end{aligned} \tag{10}$$

By substituting (10) to the right-hand part of Eqs. (9) proportional to $\beta m$, we obtain integro-differential equations for $\psi_R$ and $\psi_L$

$$\begin{aligned} &\left[\left(p_0-\boldsymbol{\alpha}\mathbf{p}-q\left(B^0-\boldsymbol{\alpha}\mathbf{B}\right)\right)-\beta m\left(p_0-\boldsymbol{\alpha}\mathbf{p}-q\left(B^0-\boldsymbol{\alpha}\mathbf{B}\right)\right)^{-1}\beta m\right]\psi_L=0 \\ &\left[\left(p_0-\boldsymbol{\alpha}\mathbf{p}-q\left(B^0-\boldsymbol{\alpha}\mathbf{B}\right)\right)-\beta m\left(p_0-\boldsymbol{\alpha}\mathbf{p}-q\left(B^0-\boldsymbol{\alpha}\mathbf{B}\right)\right)^{-1}\beta m\right]\psi_R=0 \end{aligned} \qquad (11)$$

One can see that equations for $\psi_R$ and $\psi_L$ have the same form, and, in contrast to Eqs. (9), the presence of mass $m$ does not lead to mixing the right and left components of $\psi$.

Eqs. (11) can be written as

$$\left[\left(p_0-\boldsymbol{\alpha}\mathbf{p}-q\alpha_\mu B^\mu\right)-\left(p_0+\boldsymbol{\alpha}\mathbf{p}-q\bar{\alpha}_\mu B^\mu\right)^{-1}m^2\right]\psi_{L,R}=0 \qquad (12)$$

In expression (12), designation $\psi_{L,R}$ shows that equations for $\psi_L$ and $\psi_R$ have the same form; $\bar{\alpha}_\mu=\begin{cases}1\\-\alpha^i\end{cases}$.

If we multiply Eqs. (12) on the left by the term $p_0+\boldsymbol{\alpha}\mathbf{p}-q\bar{\alpha}_\mu B^\mu$, we obtain second-order equations with respect to $p^\mu$

$$\left[\left(p_0+\boldsymbol{\alpha}\mathbf{p}-q\bar{\alpha}_\mu B^\mu\right)\left(p_0-\boldsymbol{\alpha}\mathbf{p}-q\alpha_\mu B^\mu\right)-m^2\right]\psi_{L,R}=0 \qquad (13)$$

For the case of quantum electrodynamics, $\left(q=e,\, B^\mu=A^\mu\right)$, Eqs. (13) take the form of

$$\left[\left(p_0-eA_0\right)^2-\left(\mathbf{p}-e\mathbf{A}\right)^2-m^2+e\boldsymbol{\Sigma}\mathbf{H}+ie\boldsymbol{\alpha}\mathbf{E}\right]\psi_{L,R}=0 \qquad (14)$$

In Eqs. (14) $\mathbf{H}=rot\,\mathbf{A}$ – is magnetic field, and $\mathbf{E}=-\frac{\partial\mathbf{A}}{\partial t}-\nabla A_0$ is electrical field.

Eqs. (14) coincide with the second-order equation obtained by Dirac in the 1920s for the bispinor $\psi$ [8]. However, in contrast to [8] (see also [9]), Eqs. (14) are written for the left and right components of the bispinor $\psi$ and contain no “excess" solutions. The operator $\gamma_5$ commutes with Eqs. (13), (14). Consequently, $\gamma_5\psi=\delta\psi\ \left(\delta^2=1;\delta=\pm1\right)$. The case of $\delta=-1$ corresponds to the solutions of Eqs. (13), (14) for $\psi_L$, and $\delta=+1$ corresponds to the solutions of Eqs. (13), (14) for $\psi_R$.

Thus, expressions (12), (13) show that equations of motion for initially massive fermions interacting with boson fields can be written in the chirally symmetric form.

### 3.2 Isotopic Foldy-Wouthuysen representation for massive fermions

Eqs. (12), (13) are nonlinear with respect to the operator $p_0 = i\frac{\partial}{\partial t}$.

The linear form of the chirally symmetric equations of fermion fields with respect to $p_0$ can be derived using the Foldy-Wouthuysen transformation [1] in a specially introduced isotopic space.

Let us introduce an eight-component field operator, $\Phi_1 = \begin{pmatrix} \psi_R \\ \psi_L \end{pmatrix}$ and isotopic matrices, $\tau_3 = \begin{pmatrix} E_{4\times4} & 0 \\ 0 & -E_{4\times4} \end{pmatrix}, \tau_1 = \begin{pmatrix} 0 & E_{4\times4} \\ E_{4\times4} & 0 \end{pmatrix}$ affecting the four upper and four lower components of the operator $\Phi_1$. Now, Eqs. (9) can be written as

$$p_0\Phi_1 = \left(\boldsymbol{\alpha}\mathbf{p} + \tau_1\beta m + q\alpha_\mu B^\mu\right)\Phi_1 \tag{15}$$

Since $\tau_1$ commutes with the right-hand part of Eq. (15), the field $\Phi_2 = \tau_1\Phi_1 = \begin{pmatrix} \psi_L \\ \psi_R \end{pmatrix}$ is also solution to Eq. (15).

$$p_0\Phi_2 = \left(\boldsymbol{\alpha}\mathbf{p} + \tau_1\beta m + q\alpha_\mu B^\mu\right)\Phi_2. \tag{16}$$

Finally, let us introduce a sixteen-component spinor,

$$\Phi = \begin{pmatrix} \Phi_1 \\ \Phi_2 \end{pmatrix} = \begin{pmatrix} \psi_R \\ \psi_L \\ \psi_L \\ \psi_R \end{pmatrix} \tag{17}$$

and generalization of the isotopic matrices $\tau_1$, $\tau_3$:

$$E_{16\times16}\cdot\tau_1 = \Sigma_1^I;\ E_{16\times16}\cdot\tau_3 = \Sigma_3^I;\ \rho_1^I\cdot\tau_1 = \alpha_1^I; \tag{18}$$

$$\rho_1^I = \begin{pmatrix} 0 & E_{8\times8} \\ E_{8\times8} & 0 \end{pmatrix};\ \rho_3^I = \begin{pmatrix} E_{8\times8} & 0 \\ 0 & -E_{8\times8} \end{pmatrix};\ \rho_2^I = i\rho_3^I\rho_1^I.$$

In expressions (18), the symbol "$I$" shows that the introduced matrices, similarly to the matrices $\tau_1$, $\tau_3$, act in the isotopic space without affecting the internal structure of the fields $\psi_R, \psi_L$.

The Dirac equation for $\Phi(x)$ accounting for (18) can be written as

$$p_0\Phi(x) = \left(\boldsymbol{\alpha}\mathbf{p} + \Sigma_1^I\beta m + \frac{1}{2}\left(E_{16\times16} + \alpha_1^I\right)q\alpha_\mu\beta^\mu\right)\Phi(x) \tag{19}$$

Taking into account the spinor structure of $\Phi(x)$ (see (17)), Eq. (19) contains the doubled Eqs. (9).

Eqs. (15), (16), (19) in the Dirac representation are equivalent to each other in their physical consequences; however, as shown below, each of them has its own features in the isotopic Foldy-Wouthuysen (IFW) representation.

Further, let us consider Eqs. (15) – (16) without boson fields $B^{\mu}$ (free motion):

$$p_0 \Phi_{1,2} = \left( \boldsymbol{\alpha}\mathbf{p} + \tau_1 \beta m \right) \Phi_{1,2} \tag{20}$$

$\Phi_{1,2}$ shows that Eqs. (20) are the same for the fields $\Phi_1$, $\Phi_2$.

Let us find the Foldy-Wouthuysen transformation in the isotopic space for Eqs. (20) using the Eriksen transformation [3].

$$\left(U_0\right)_{IFW} = U_{Er} = \frac{1}{2}\left(1+\tau_3\lambda\right)\left(\frac{1}{2}+\frac{\tau_3\lambda+\lambda\tau_3}{4}\right)^{-1/2} \tag{21}$$

In expression (21), $\lambda = \dfrac{\boldsymbol{\alpha}\mathbf{p}+\tau_1\beta m}{E}$; $E = \left(\mathbf{p}^2+m^2\right)^{1/2}$. Since $\left(\boldsymbol{\alpha}\mathbf{p}+\tau_1\beta m\right)^2 = E^2$, $\lambda^2 = 1$.

Expression (21) can be transformed as follows:

$$\begin{aligned}\left(U_0\right)_{IFW} &= U_{Er} = \frac{1}{2}\left(1+\frac{\tau_3\boldsymbol{\alpha}\mathbf{p}+\tau_3\tau_1\beta m}{E}\right)\left(\frac{1}{2}+\frac{\tau_3\boldsymbol{\alpha}\mathbf{p}}{2E}\right)^{-1/2} = \\ &= \sqrt{\frac{E+\tau_3\boldsymbol{\alpha}\mathbf{p}}{2E}}\left(1+\frac{1}{E+\tau_3\boldsymbol{\alpha}\mathbf{p}}\tau_3\tau_1\beta m\right)\end{aligned} \tag{22}$$

Transformation (22) is unitary $\left(\left(U_0\right)_{IFW}\left(U_0\right)^{\dagger}_{IFW} = 1\right)$, and

$$\left(H_0\right)_{IFW} = \left(U_0\right)_{IFW}\left(\boldsymbol{\alpha}\mathbf{p}+\tau_1\beta m\right)\left(U_0\right)^{\dagger}_{IFW} = \tau_3 E \tag{23}$$

Thus, Eqs. (20) in the isotopic Foldy-Wouthuysen representation have the form

$$p_0\left(\Phi_{1,2}\right)_{IFW} = \tau_3 E\left(\Phi_{1,2}\right)_{IFW} \tag{24}$$

Since $\left(U_0\right)_{IFW}\tau_1\left(U_0\right)^{\dagger}_{IFW} = \tau_3\beta$, a analog of the relationship $\Phi_2 = \tau_1\Phi_1$ in the *IFW* representation is the relationship

$$\left(\Phi_2\right)_{IFW} = \tau_3\beta\left(\Phi_1\right)_{IFW}. \tag{25}$$

At transition to the Foldy-Wouthuysen representation, in addition to the condition that the Hamiltonian should be block-diagonal in Eq. (24), the upper or lower components of $\Phi_1, \Phi_2$ should necessarily be zero (condition of reduction of the fields $\Phi_1, \Phi_2$) [6].

Let us check whether this condition is met in our case. Taking into account Eqs. (9), (10), the normalized solutions to Eqs. (20) for the wave functions $\Phi_1, \Phi_2$ can be expressed as follows:

$$\Phi_1^{(+)}(\mathbf{x},t)=e^{-iEt}\sqrt{\frac{E+\mathbf{\Sigma p}}{2E}}\begin{pmatrix}\psi_R^{(+)}(\mathbf{x})\\ \frac{1}{E-\boldsymbol{\alpha}\mathbf{p}}\beta m\psi_R^{(+)}(\mathbf{x})\end{pmatrix};\ \Phi_1^{(-)}(\mathbf{x},t)=e^{iEt}\sqrt{\frac{E+\mathbf{\Sigma p}}{2E}}\begin{pmatrix}-\frac{1}{E+\boldsymbol{\alpha}\mathbf{p}}\beta m\psi_L^{(-)}(\mathbf{x})\\ \psi_L^{(-)}(\mathbf{x})\end{pmatrix}$$

$$\Phi_2^{(+)}(\mathbf{x},t)=e^{-iEt}\sqrt{\frac{E-\mathbf{\Sigma p}}{2E}}\begin{pmatrix}\psi_L^{(+)}(\mathbf{x})\\ \frac{1}{E-\boldsymbol{\alpha}\mathbf{p}}\beta m\psi_L^{(+)}(\mathbf{x})\end{pmatrix};\ \Phi_2^{(-)}(\mathbf{x},t)=e^{iEt}\sqrt{\frac{E-\mathbf{\Sigma p}}{2E}}\begin{pmatrix}-\frac{1}{E+\boldsymbol{\alpha}\mathbf{p}}\beta m\psi_R^{(-)}(\mathbf{x})\\ \psi_R^{(-)}(\mathbf{x})\end{pmatrix} \tag{26}$$

In Eqs. (26), $\Phi_1^{(+)},\Phi_2^{(+)};\Phi_1^{(-)},\Phi_2^{(-)}$ are solutions to Eqs. (20) with positive and negative energy, respectively.

If the matrices $\alpha^i$, $\beta$ are used in the Dirac-Pauli representation, right and left components of the wave functions for solutions with positive and negative energy are equal to

$$\psi_R^{(+)}(\mathbf{x})=\frac{1}{2}(1+\gamma_5)\psi_D^{(+)}(\mathbf{x})=\frac{1}{2}\sqrt{\frac{E+m}{2E}}\begin{pmatrix}\left(1+\frac{\boldsymbol{\sigma}\mathbf{p}}{E+m}\right)\varphi^{(+)}(\mathbf{x})\\ \left(1+\frac{\boldsymbol{\sigma}\mathbf{p}}{E+m}\right)\varphi^{(+)}(\mathbf{x})\end{pmatrix}$$

$$\psi_R^{(-)}(\mathbf{x})=\frac{1}{2}(1+\gamma_5)\psi_D^{(-)}(\mathbf{x})=\frac{1}{2}\sqrt{\frac{E+m}{2E}}\begin{pmatrix}\left(1-\frac{\boldsymbol{\sigma}\mathbf{p}}{E+m}\right)\chi^{(-)}(\mathbf{x})\\ \left(1-\frac{\boldsymbol{\sigma}\mathbf{p}}{E+m}\right)\chi^{(-)}(\mathbf{x})\end{pmatrix}$$

$$\psi_L^{(+)}(\mathbf{x})=\frac{1}{2}(1-\gamma_5)\psi_D^{(+)}(\mathbf{x})=\frac{1}{2}\sqrt{\frac{E+m}{2E}}\begin{pmatrix}\left(1-\frac{\boldsymbol{\sigma}\mathbf{p}}{E+m}\right)\varphi^{(+)}(\mathbf{x})\\ -\left(1-\frac{\boldsymbol{\sigma}\mathbf{p}}{E+m}\right)\varphi^{(+)}(\mathbf{x})\end{pmatrix} \tag{27}$$

$$\psi_L^{(-)}(\mathbf{x})=\frac{1}{2}(1-\gamma_5)\psi_D^{(-)}(\mathbf{x})=\frac{1}{2}\sqrt{\frac{E+m}{2E}}\begin{pmatrix}-\left(1+\frac{\boldsymbol{\sigma}\mathbf{p}}{E+m}\right)\chi^{(-)}(\mathbf{x})\\ \left(1+\frac{\boldsymbol{\sigma}\mathbf{p}}{E+m}\right)\chi^{(-)}(\mathbf{x})\end{pmatrix}$$

In expressions (27), $\varphi^{(+)}(\mathbf{x}),\chi^{(-)}(\mathbf{x})$ are normalized two-component solutions of the Dirac equation with positive and negative energy. In Eqs. (26), (27), $E$ and $\mathbf{p}$ are energy and momentum operators of a Dirac particle.

According to Eqs. (27),

$$\begin{gathered}\boldsymbol{\alpha}\mathbf{p}\,\psi_R^{(+)}(\mathbf{x})=\mathbf{\Sigma p}\,\psi_R^{(+)}(\mathbf{x});\ \boldsymbol{\alpha}\mathbf{p}\,\psi_R^{(-)}(\mathbf{x})=\mathbf{\Sigma p}\,\psi_R^{(-)}(\mathbf{x})\\ \boldsymbol{\alpha}\mathbf{p}\,\psi_L^{(+)}(\mathbf{x})=-\mathbf{\Sigma p}\,\psi_L^{(+)}(\mathbf{x});\ \boldsymbol{\alpha}\mathbf{p}\,\psi_L^{(-)}(\mathbf{x})=-\mathbf{\Sigma p}\,\psi_L^{(-)}(\mathbf{x})\end{gathered} \tag{28}$$

Eqs. (27) lead to the following normalizing conditions:

$$\begin{aligned}
\psi_R^{(+)\dagger}(\mathbf{x})\psi_R^{(+)}(\mathbf{x}) &= \varphi^{(+)\dagger}(\mathbf{x})\frac{E+\boldsymbol{\sigma}\mathbf{p}}{2E}\varphi^{(+)}(\mathbf{x}) \\
\psi_R^{(-)\dagger}(\mathbf{x})\psi_R^{(-)}(\mathbf{x}) &= \chi^{(-)\dagger}(\mathbf{x})\frac{E-\boldsymbol{\sigma}\mathbf{p}}{2E}\chi^{(-)}(\mathbf{x}) \\
\psi_L^{(+)\dagger}(\mathbf{x})\psi_L^{(+)}(\mathbf{x}) &= \varphi^{(+)\dagger}(\mathbf{x})\frac{E-\boldsymbol{\sigma}\mathbf{p}}{2E}\varphi^{(+)}(\mathbf{x}) \\
\psi_L^{(-)\dagger}(\mathbf{x})\psi_L^{(-)}(\mathbf{x}) &= \chi^{(-)\dagger}(\mathbf{x})\frac{E+\boldsymbol{\sigma}\mathbf{p}}{2E}\chi^{(-)}(\mathbf{x}).
\end{aligned} \tag{29}$$

By applying the transformation matrix $(U_0)_{IFW}$ (22) to $\Phi_1, \Phi_2$ (see (26)) we obtain

$$\begin{aligned}
\Phi_{1IFW}^{(+)}(\mathbf{x},t) &= (U_0)_{IFW}\,\Phi_1^{(+)}(\mathbf{x},t) = e^{-iEt}\begin{pmatrix} \sqrt{\dfrac{2E}{E+\boldsymbol{\Sigma}\mathbf{p}}}\psi_R^{(+)}(\mathbf{x}) \\ 0 \end{pmatrix} = e^{-iEt}\frac{1}{\sqrt{2}}\begin{pmatrix} \varphi^{(+)}(\mathbf{x}) \\ \varphi^{(+)}(\mathbf{x}) \\ 0 \\ 0 \end{pmatrix}; \\
\Phi_{1IFW}^{(-)}(\mathbf{x},t) &= (U_0)_{IFW}\,\Phi_1^{(-)}(\mathbf{x},t) = e^{iEt}\begin{pmatrix} 0 \\ \sqrt{\dfrac{2E}{E+\boldsymbol{\Sigma}\mathbf{p}}}\psi_L^{(-)}(\mathbf{x}) \end{pmatrix} = e^{iEt}\frac{1}{\sqrt{2}}\begin{pmatrix} 0 \\ 0 \\ -\chi^{(-)}(\mathbf{x}) \\ \chi^{(-)}(\mathbf{x}) \end{pmatrix}; \\
\Phi_{2IFW}^{(+)}(\mathbf{x},t) &= (U_0)_{IFW}\,\Phi_2^{(+)}(\mathbf{x},t) = e^{-iEt}\begin{pmatrix} \sqrt{\dfrac{2E}{E-\boldsymbol{\Sigma}\mathbf{p}}}\psi_L^{(+)}(\mathbf{x}) \\ 0 \end{pmatrix} = e^{-iEt}\frac{1}{\sqrt{2}}\begin{pmatrix} \varphi^{(+)}(\mathbf{x}) \\ -\varphi^{(+)}(\mathbf{x}) \\ 0 \\ 0 \end{pmatrix}; \\
\Phi_{2IFW}^{(-)}(\mathbf{x},t) &= (U_0)_{IFW}\,\Phi_2^{(-)}(\mathbf{x},t) = e^{iEt}\begin{pmatrix} 0 \\ \sqrt{\dfrac{2E}{E-\boldsymbol{\Sigma}\mathbf{p}}}\psi_R^{(-)}(\mathbf{x}) \end{pmatrix} = e^{iEt}\frac{1}{\sqrt{2}}\begin{pmatrix} 0 \\ 0 \\ \chi^{(-)}(\mathbf{x}) \\ \chi^{(-)}(\mathbf{x}) \end{pmatrix}.
\end{aligned} \tag{30}$$

One can see from relationships (30) that the reduction condition is fulfilled and the matrix $(U_0)_{IFW}$ is, indeed, the Foldy-Wouthuysen tranformation for the fields $\Phi_1, \Phi_2$ in the introduced isotopic space.

For the Dirac matrices in the chiral representation, following (27) – (30), one can obtain the following form of the basis functions in the IFW representation:

$$\Phi^{(+)}_{1IFW}(\mathbf{x},t) = e^{-iEt}\begin{pmatrix}\sqrt{\dfrac{2E}{E+\mathbf{\Sigma p}}}\psi^{(+)}_R(\mathbf{x}) \\ 0\end{pmatrix} = e^{-iEt}\begin{pmatrix}\varphi^{(+)}(\mathbf{x}) \\ 0 \\ 0 \\ 0\end{pmatrix};$$

$$\Phi^{(-)}_{1IFW}(\mathbf{x},t) = e^{iEt}\begin{pmatrix}\sqrt{\dfrac{2E}{E+\mathbf{\Sigma p}}}\psi^{(+)}_R(\mathbf{x}) \\ 0\end{pmatrix} = e^{iEt}\begin{pmatrix}0 \\ 0 \\ 0 \\ \chi^{(-)}(\mathbf{x})\end{pmatrix};$$

$$\Phi^{(+)}_{2IFW}(\mathbf{x},t) = e^{-iEt}\begin{pmatrix}\sqrt{\dfrac{2E}{E+\mathbf{\Sigma p}}}\psi^{(+)}_R(\mathbf{x}) \\ 0\end{pmatrix} = e^{-iEt}\begin{pmatrix}0 \\ \varphi^{(+)}(\mathbf{x}) \\ 0 \\ 0\end{pmatrix}; \qquad (31)$$

$$\Phi^{(-)}_{2IFW}(\mathbf{x},t) = e^{iEt}\begin{pmatrix}\sqrt{\dfrac{2E}{E+\mathbf{\Sigma p}}}\psi^{(+)}_R(\mathbf{x}) \\ 0\end{pmatrix} = e^{-iEt}\begin{pmatrix}0 \\ 0 \\ -\chi^{(-)}(\mathbf{x}) \\ 0\end{pmatrix}.$$

Naturally, in the chiral representation relationship (25) that links $\Phi_{1IFW}(x)$ with $\Phi_{2IFW}(x)$ remains valid.

Basis functions (30), (31) are also solutions to free equation (19) transformed to the IFW representation. Indeed, taking into account definitions (17), (18), Eq. (19) for free motion has the form

$$p_0\begin{pmatrix}\Phi_1(x) \\ \Phi_2(x)\end{pmatrix} = (\boldsymbol{\alpha}\mathbf{p} + E_{16\times16}\tau_1\beta m)\begin{pmatrix}\Phi_1(x) \\ \Phi_2(x)\end{pmatrix}. \qquad (32)$$

By applying transformation matrix (22) to the upper and lower components of Eq. (32), we obtain

$$p_0\begin{pmatrix}\Phi_{1IFW}(x) \\ \Phi_{2IFW}(x)\end{pmatrix} = E_{16\times16}\tau_3 E\begin{pmatrix}\Phi_{1IFW}(x) \\ \Phi_{2IFW}(x)\end{pmatrix}. \qquad (33)$$

Basis functions (30), (31) are orthonormalized and complete. Completeness conditions include both functions $\Phi^{(\pm)}_{1IFW}$ and $\Phi^{(\pm)}_{2IFW}$.

$$\sum_{\pm s}\Phi^{(+)}_{1IFW}(\mathbf{x},t)_\alpha \Phi^{(+)\dagger}_{1IFW}(\mathbf{x}',t)_\beta = \left[\frac{1}{2}(1+\tau_3)\frac{1}{2}(1+\gamma^5)\right]_{\alpha\beta}\delta(\mathbf{x}-\mathbf{x}') \qquad (34)$$

$$\sum_{\pm s}\Phi^{(+)}_{2IFW}(\mathbf{x},t)_\alpha \Phi^{(+)\dagger}_{2IFW}(\mathbf{x}',t)_\beta = \left[\frac{1}{2}(1+\tau_3)\frac{1}{2}(1-\gamma^5)\right]_{\alpha\beta}\delta(\mathbf{x}-\mathbf{x}') \qquad (35)$$

$$\sum_{\pm s}\Phi^{(-)}_{1IFW}(\mathbf{x},t)_\alpha \Phi^{(-)\dagger}_{1IFW}(\mathbf{x}',t)_\beta = \left[\frac{1}{2}(1-\tau_3)\frac{1}{2}(1-\gamma^5)\right]_{\alpha\beta}\delta(\mathbf{x}-\mathbf{x}') \qquad (36)$$

$$\sum_{\pm s}\Phi^{(-)}_{2IFW}(\mathbf{x},t)_\alpha\,\Phi^{(-)\dagger}_{2IFW}(\mathbf{x}',t)_\beta=\left[\frac{1}{2}\left(1-\tau_3\right)\frac{1}{2}\left(1+\gamma^5\right)\right]_{\alpha\beta}\delta(\mathbf{x}-\mathbf{x}') \tag{37}$$

$$\begin{aligned}&\sum_{\pm s}(\Phi^{(+)}_{1IFW}(\mathbf{x},t)_\alpha\,\Phi^{(+)}_{1IFW}(\mathbf{x}',t)_\beta+\Phi^{(-)}_{1IFW}(\mathbf{x},t)_\alpha\,\Phi^{(-)}_{1IFW}(\mathbf{x}',t)_\beta+\\&+\Phi^{(+)}_{2IFW}(\mathbf{x},t)_\alpha\,\Phi^{(+)\dagger}_{2IFW}(\mathbf{x}',t)_\beta+\Phi^{(-)}_{2IFW}(\mathbf{x},t)_\alpha\,\Phi^{(-)}_{2IFW}(\mathbf{x}',t)_\beta)=\delta_{\alpha\beta}\delta(\mathbf{x}-\mathbf{x}').\end{aligned} \tag{38}$$

Let us also note the following useful relationships: expressions $\sum_{\pm s}\Phi^{(+)}_{1IFW}\left(\mathbf{x},t\right)_\alpha\,\Phi^{(+)}_{2IFW}\left(\mathbf{x}',t\right)_\beta$ and $\sum_{\pm s}\Phi^{(+)}_{2IFW}\left(\mathbf{x},t\right)_\alpha\,\Phi^{(+)}_{1IFW}\left(\mathbf{x}',t\right)_\beta$ differ from (34) and (35), respectively, in the presence of the matrix $\beta$ after the square bracket; expressions $\sum_{\pm s}\Phi^{(-)}_{1IFW}\left(\mathbf{x},t\right)_\alpha\,\Phi^{(-)}_{2IFW}\left(\mathbf{x}',t\right)_\beta$ and $\sum_{\pm s}\Phi^{(-)}_{2IFW}\left(\mathbf{x},t\right)_\alpha\,\Phi^{(-)}_{1IFW}\left(\mathbf{x}',t\right)_\beta$ differ from (36) and (37), respectively, in the presence of the matrix $\left(-\beta\right)$ after the square bracket.

In the presence of boson fields $B^\mu(x)$ interacting with fermion fields $\Phi_1(x),\Phi_2(x),$ the isotopic Foldy-Wouthuysen transformation and the form of the Hamiltonians of Eqs. (15), (16), (19) in the IFW representation can be obtained as a series in powers of the coupling constant using the algorithm described in Refs. [4], [5].

As a result, using the notation from Refs. [4], [5], we obtain

$$\begin{aligned}&\left(\Phi_{1,2}\right)_{IFW}=U_{IFW}\Phi_{1,2};\\&U_{IFW}=\left(1+\delta_1+\delta_2+\delta_3+...\right)(U_0)_{IFW};\end{aligned} \tag{39}$$

$$\begin{aligned}&\left(\Phi\right)_{IFW}=U^{\Phi}_{IFW}\Phi;\\&U^{\Phi}_{IFW}=\left(1+\delta^{\Phi}_1+\delta^{\Phi}_2+\delta^{\Phi}_3+...\right)(U^{\Phi}_0)_{IFW}.\end{aligned} \tag{40}$$

By applying the same transformation (39) to each of Eqs. (15), (16), we obtain the following equations and Hamiltonian densities in the isotopic Foldy-Wouthuysen representation:

$$p_0\Phi_{IFW}=\left(\tau_3E+K_1+K_2+K_3+...\right)(\Phi_1)_{IFW}; \tag{41}$$

$$\mathscr{H}^{I}_{IFW}=(\Phi_1)^\dagger_{IFW}\left(\tau_3E+K_1+K_2+K_3+...\right)(\Phi_1)_{IFW}; \tag{42}$$

$$p_0(\Phi_2)_{IFW}=\left(\tau_3E+K_1+K_2+K_3+...\right)(\Phi_2)_{IFW}; \tag{43}$$

$$\mathscr{H}^{II}_{IFW}=(\Phi_2)^\dagger_{IFW}\left(\tau_3E+K_1+K_2+K_3+...\right)(\Phi_2)_{IFW}. \tag{44}$$

Similarly, by applying transformation (40) to Eq. (19), we obtain

$$p_0\Phi_{IFW}=\left(E_{16\times16}\tau_3E+K^{\Phi}_1+K^{\Phi}_2+K^{\Phi}_3+...\right)\Phi_{IFW}; \tag{45}$$

$$\mathscr{H}^{III}_{IFW}=(\Phi)^\dagger_{IFW}\left(E_{16\times16}\tau_3E+K^{\Phi}_1+K^{\Phi}_2+K^{\Phi}_3+...\right)(\Phi)_{IFW}. \tag{46}$$

For illustration, explicit expressions in the momentum representation for eight-dimensional operators $K_1$, $K_2$, $K_3$ in (41) – (44) are presented in Appendix 1. Sixteen-dimensional operators $K_n^{\Phi}$ in (45), (46) constitute generalized expressions $K_n$ with the following replacements:

$$\tau_1 \to E_{16\times16}\tau_1,\ \tau_3 \to E_{16\times16}\tau_3,\ q \to q^{\Phi} = \frac{1}{2}\left(E_{16\times16} + \alpha_1^I\right)q \qquad (47)$$

Expressions for the operators $C$, $N$, $C^{\Phi}$, $N^{\Phi}$, which, in accordance with [4], [5], form the basis for writing the interaction Hamiltonian in the FW representation, are written for the isotopic Foldy-Wouthuysen representation as follows:

$$C = \left[\left(U_0\right)_{IFW} q\alpha_\mu B_\mu \left(U_0\right)_{IFW}^{\dagger}\right]^{even} = R\left(qB^0 - LqB^0L\right)R - R\left(q\boldsymbol{\alpha}\mathbf{B} - Lq\boldsymbol{\alpha}\mathbf{B}L\right)R$$

$$N = \left[\left(U_0\right)_{IFW} q\alpha_\mu B_\mu \left(U_0\right)_{IFW}^{\dagger}\right]^{odd} = R\left(LqB^0 - qB^0L\right)R - R\left(Lq\boldsymbol{\alpha}\mathbf{B} - q\boldsymbol{\alpha}\mathbf{B}L\right)R \qquad (48)$$

$$R = \sqrt{\frac{E + \tau_3\boldsymbol{\alpha}\mathbf{p}}{2E}};\ L = \frac{1}{E + \tau_3\boldsymbol{\alpha}\mathbf{p}}\tau_3\tau_1\beta m$$

Expressions for the sixteen-dimensional operators $C^{\Phi}$, $N^{\Phi}$, in (45), (46) are derived from expressions (48) with replacement (47):

$$C^{\Phi} = R^{\Phi}\left(q^{\Phi}B^0 - L^{\Phi}q^{\Phi}B^0L^{\Phi}\right)R^{\Phi} - R^{\Phi}\left(q^{\Phi}\boldsymbol{\alpha}\mathbf{B} - L^{\Phi}q^{\Phi}\boldsymbol{\alpha}\mathbf{B}L^{\Phi}\right)R^{\Phi}$$

$$N^{\Phi} = R^{\Phi}\left(L^{\Phi}q^{\Phi}B^0 - q^{\Phi}B^0L^{\Phi}\right)R^{\Phi} - R^{\Phi}\left(L^{\Phi}q^{\Phi}\boldsymbol{\alpha}\mathbf{B} - q^{\Phi}\boldsymbol{\alpha}\mathbf{B}L^{\Phi}\right)R^{\Phi} \qquad (49)$$

$$R^{\Phi} = \sqrt{\frac{E + E_{16\times16}\tau_3\boldsymbol{\alpha}\mathbf{p}}{2E}};\ L^{\Phi} = \frac{1}{E + E_{16\times16}\tau_3\boldsymbol{\alpha}\mathbf{p}}E_{16\times16}\tau_3E_{16\times16}\tau_1\beta m$$

The superscripts $even, odd$ in Eqs. (48) show the even and odd parts of the operators, which respectively do not link and link the upper and lower isotopic components of $(\Phi_1)_{IFW}$ and $(\Phi_2)_{IFW}$.

Note that when solving applied problems in the quantum field theory using the perturbation theory, fermion fields are expanded in solutions of Dirac equations for free motion or for motion in static external fields. In our case, in the isotopic Foldy-Wouthuysen representation, we can also expand the corresponding fermion fields in the basis of solutions of Eqs. (30), (31) or in a similar basis of solutions of the isotopic Foldy-Wouthuysen equations in static external fields and express Hamiltonians (42), (44), (46) in their terms.

By definition, the eight-dimensional operators $K_n$ in Hamiltonians (42), (44) are even with respect to the upper and lower isotopic components and therefore contain an even number of odd operators . The sixteen-dimensional operators $K^{\Phi}$ contain even and odd parts because of the form of $q^{\Phi}$ in (47). In their structure, the operators $K_n^{\Phi}$ are similar to the operators $K_n$, so they also contain an even number of operators $N^{\Phi}$. The odd parts of the operators $K^{\Phi}$ make it possible to link the fields $\Phi_{1FW}$, $\Phi_{2FW}$ and, taking into account their structure, provide for the interaction only between left particles and left antiparticles, and only between right particles and right antiparticles.

Thus, the analysis shows that the equations of fermion fields (41), (43), (45) and their Hamiltonians in the isotopic Foldy-Wouthuysen representation for the case of the considered interaction are written in the chirally symmetric form irrespective of whether fermions possess mass or are massless.

### 3.3 Chiral invariance in the isotopic Foldy-Wouthuysen representation

In the isotopic Foldy-Wouthuysen representation for free motion, Hamiltonian (23) and Eqs. (24) are invariant with respect to the chiral transformation

$$\left(\Phi_{1,2}\right)_{IFW} \to e^{i\alpha\gamma^5}\left(\Phi_{1,2}\right)_{IFW} \tag{50}$$

This invariance is independent of whether fermions possess mass or not.

Note that if we use the operator

$$\left(\gamma^5\right)_{IFW} = \left(U_0\right)_{IFW}\gamma^5\left(U_0\right)^{\dagger}_{IFW} = \gamma^5\left(\frac{\tau_3\boldsymbol{\alpha}\mathbf{p}}{E} - \frac{\tau_3\tau_1\beta m}{E}\right) \tag{51}$$

in the indices of exponent in expression (50), then, for a nonzero fermion mass, Hamiltonian (23) and Eqs. (24) are non-invariant with respect to the transformation in the IFW representation

$$\left(\Phi_{1,2}\right)_{IFW} \to e^{i\alpha(\gamma^5)_{IFW}}\left(\Phi_{1,2}\right)_{IFW}, \tag{52}$$

like in the case of a similar transformation with the operator $\gamma^5$ in the Dirac representation for massive fermions

$$\left(\Phi_{1,2}\right) \to e^{i\alpha(\gamma^5)}\left(\Phi_{1,2}\right). \tag{53}$$

The operator $(\gamma^5)_{IFW}$ is associated with fermion helicity in the isotopic Foldy-Wouthuysen representation.

Since the basis functions of free motion (30), (31) in the IFW representation preserve the form of the right and left components of the field operators used in the Dirac representation $\left(\psi_{R,L} = \frac{1}{2}\left(1\pm\gamma^5\right)\psi\right)$, the choice of chiral transformation (50) is quite justified.

For the interaction of $q\alpha_\mu B^\mu(x)$ with the matrices $\alpha_\mu$ commuting with $\gamma^5$, as chosen by us in the Dirac representation, chiral symmetry of equations and Hamiltonians (41) – (46) is preserved, as well. Indeed, in expressions (41) – (46), the only operators that do not commute with $\gamma^5$ are the operators $L$, $L^{\Phi}$ (see (48), (49)). However, in the expressions for $C$ and $C^{\Phi}$ (see (48), (49)), the operators $L$, $L^{\Phi}$ are contained in the even form. Expressions $N$ and $N^{\Phi}$ in the even form are also contained in the operators $K_n$ and $K_n^{\Phi}$. Consequently, the operator $\gamma^5$ commutes with the Hamiltonians of Eqs. (41), (43), (45).

This analysis shows that the equations of massive fermion fields (41), (43), (45) and their Hamiltonians in the isotopic Foldy-Wouthuysen representation are invariant with respect to chiral transformations (50).

Accordingly, because of the unitary property of the IFW transformation, equivalent Dirac equations (15), (16), (19) with the interaction $q\alpha_{\mu}B^{\mu}(x)$ are invariant with respect to the transformations

$$\Phi_{1,2} \rightarrow e^{i\alpha(\gamma^5)_D}\Phi_{1,2} \tag{54}$$

$$\left(\gamma^5\right)_D = U_{IFW}^{\dagger}\gamma^5 U_{IFW}. \tag{55}$$

For the free case, expression (55) equals

$$\left(\gamma^5\right)_D = \left(U_0\right)_{FW}^{\dagger}\gamma^5\left(U_0\right)_{FW} = \gamma^5\left(\frac{\tau_3\boldsymbol{\alpha}\mathbf{p}}{E} + \frac{\tau_3\tau_1\beta m}{E}\right). \tag{56}$$

In equality (56), the operator $E = \left(m^2 + \mathbf{p}^2\right)^{1/2}$ is non-local with an infinite number of differentiation operators $p^i = -i\dfrac{\partial}{\partial x^i}$, $i = 1,2,3$.

In case of interaction, operator (55) is a complex expression depending on the boson fields .

As far as the author knows, invariance of the Dirac equation with respect to transformation (54) has not been considered in literature before.

### 3.4 Physical content of chirally symmetric equations and their Hamiltonians in the isotopic Foldy-Wouthuysen representation

Now, let us consider physical content of Hamiltonians and Eqs. (41) – (46) by means of symbolic patterns in Figs 1, 2, 3.

Let us first consider Eqs. (45) and the Hamiltonian $\mathcal{H}_{IFW}^{III}$ (46). The corresponding physical diagram is shown symbolically in Fig. 1.

$$E>0 \qquad T_3=+\frac{1}{2}$$

$$\Phi^{(+)}_{IFW}(\mathbf{x},t)=\begin{pmatrix}\Phi^{(+)}_{1IFW}(\mathbf{x},t)\\ \Phi^{(+)}_{2IFW}(\mathbf{x},t)\end{pmatrix}=e^{-iEt}\begin{pmatrix}A\psi^{(+)}_R(\mathbf{x})\\ 0\\ A_1\psi^{(+)}_L(\mathbf{x})\\ 0\end{pmatrix}$$

$\mathcal{H}^{III}_{IFW}$

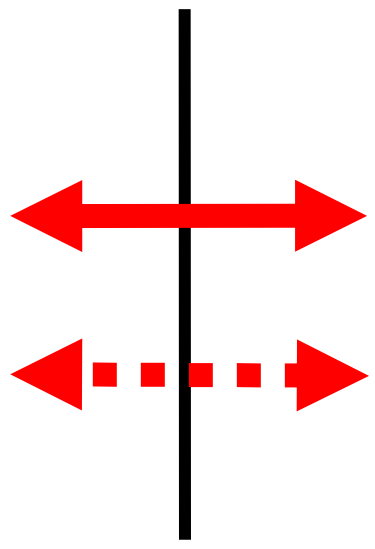

$$E<0 \qquad T_3=-\frac{1}{2}$$

$$\Phi^{(-)}_{IFW}(\mathbf{x},t)=\begin{pmatrix}\Phi^{(-)}_{1IFW}(\mathbf{x},t)\\ \Phi^{(-)}_{2IFW}(\mathbf{x},t)\end{pmatrix}=e^{iEt}\begin{pmatrix}0\\ A_1\psi^{(-)}_L(\mathbf{x})\\ 0\\ A\psi^{(-)}_R(\mathbf{x})\end{pmatrix}$$

Fig 1.

The left half-plane in Fig. 1 represents the states of basis functions (30) with the isotopic spin $T_3=+\frac{1}{2}$ and positive energy $E>0$; the right half-plane shows the states with $T_3=-\frac{1}{2}$ and negative energy $E<0$. The Hamiltonian contains the states of left and right fermions and left and right antifermions; particles and antiparticles interact with each other in a real (solid arrow in Fig 1) and virtual (dashed arrow in Fig. 1) manner. The physical diagram in Fig. 1 corresponds to the world immediately around us.

Let us proceed to considering the Hamiltonian $\mathcal{H}^{I}_{IFW}$ (42) и $\mathcal{H}^{II}_{IFW}$ (44) with respective Eqs. (41) and (43). The symbolically representation is shown in Fig. 2.3.

$$E>0 \qquad T_3=+\frac{1}{2}$$

$$\Phi^{(+)}_{IFW}(\mathbf{x},t)=e^{-iEt}\begin{pmatrix}A\psi^{(+)}_R(\mathbf{x})\\ 0\end{pmatrix}$$

$\mathcal{H}^{I}_{IFW}$

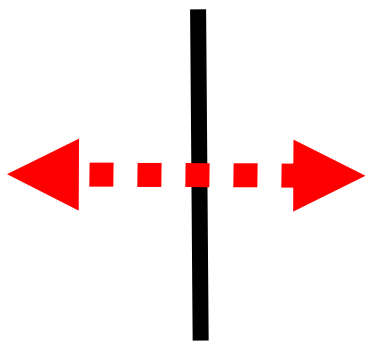

$$E<0 \qquad T_3=-\frac{1}{2}$$

$$\Phi^{(-)}_{IFW}(\mathbf{x},t)=e^{iEt}\begin{pmatrix}0\\ A_1\psi^{(-)}_L(\mathbf{x})\end{pmatrix}$$

Fig 2.

$$E>0 \qquad T_3=+\frac{1}{2}$$

$$\Phi^{(+)}_{2IFW}(\mathbf{x},t)=e^{-iEt}\begin{pmatrix}A_1\psi^{(+)}_L(\mathbf{x})\\ 0\end{pmatrix}$$

$\mathcal{H}^{II}_{IFW}$

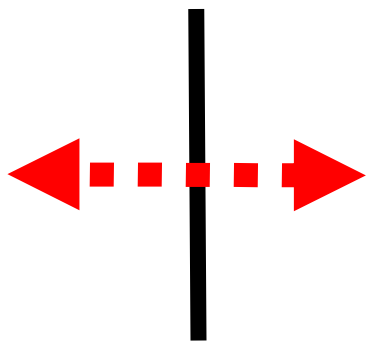

$$E<0 \qquad T_3=-\frac{1}{2}$$

$$\Phi^{(-)}_{2IFW}(\mathbf{x},t)=e^{iEt}\begin{pmatrix}0\\ A\psi^{(-)}_R(\mathbf{x})\end{pmatrix}$$

Fig 3.

It follows from Fig.2, 3 that Hamiltonians $\mathcal{H}^{I}_{IFW}$, $\mathcal{H}^{II}_{IFW}$ provide existence either right fermions and left antifermions, or left fermions and right antifermions. In both cases there is not an interaction of real particles and antiparticles.

Hamiltonian densities (42), (44), (46) and equations of fermion fields (41), (43), (45), by analogy with [4], [5], make it possible to formulate Feynman rules for calculating specific physical processes of the quantum theory of interacting fields using the methods of the perturbation theory .

## 4. Continuity equation. Conservation of vector and axial current of massive fermions in the isotopic Foldy-Wouthuysen representation

The continuity equation for free motion can be derived using Eqs. (24), (33) and Hermitian-conjugated equations.

The analysis below will be provided for Eqs. (24) with eight-component spinors $\left(\Phi_{1,2}\right)_{IFW}$. The analysis for Eqs. (33) with sixteen-component spinors $\Phi_{IFW}$ leads to the same results and conclusions.

$$\begin{aligned} i\frac{\partial\left(\Phi_{1,2}\right)_{IFW}}{\partial t} &= \tau_3 E\left(\Phi_{1,2}\right)_{IFW} \\ -i\frac{\partial\left(\Phi_{1,2}\right)^{\dagger}_{IFW}}{\partial t} &= (E\left(\Phi_{1,2}\right)^{\dagger}_{IFW})\tau_3 \end{aligned} \tag{57}$$

$$\frac{\partial}{\partial t}\left(\left(\Phi_{1,2}\right)^{\dagger}_{IFW}\left(\Phi_{1,2}\right)_{IFW}\right) = -i\left(\left(\Phi_{1,2}\right)^{\dagger}_{IFW}\tau_3 E\left(\Phi_{1,2}\right)_{IFW} - \left(E\left(\Phi_{1,2}\right)^{\dagger}_{IFW}\right)\tau_3\left(\Phi_{1,2}\right)_{IFW}\right) = -div\,\mathbf{j} \tag{58}$$

$$\begin{aligned} j^i = {} & \frac{1}{2m}\left(\left(\Phi_{1,2}\right)^{\dagger}_{IFW}p^i\tau_3(\Phi_1)_{IFW} - \left(p^i\left(\Phi_{1,2}\right)^{\dagger}_{IFW}\right)\tau_3\left(\Phi_{1,2}\right)_{IFW}\right) - \\ & -\frac{1}{8m^3}\mathbf{p}^2\left(\left(\Phi_{1,2}\right)^{\dagger}_{IFW}p^i\tau_3\left(\Phi_{1,2}\right)_{IFW} - \left(p^i\left(\Phi_{1,2}\right)^{\dagger}_{IFW}\right)\tau_3\left(\Phi_{1,2}\right)_{IFW}\right) + \\ & +\frac{1}{16m^5}\mathbf{p}^2\left(\left(\Phi_{1,2}\right)^{\dagger}_{IFW}p^i\mathbf{p}^2\tau_3\left(\Phi_{1,2}\right)_{IFW} - \left(p^i\mathbf{p}^2\left(\Phi_{1,2}\right)^{\dagger}_{IFW}\right)\tau_3\left(\Phi_{1,2}\right)_{IFW}\right) - \cdots \end{aligned} \tag{59}$$

Expression (59) for the current $j^i$ was obtained by expanding the operator $E = \sqrt{\mathbf{p}^2 + m^2}$ on the right side of (58) in powers $\frac{\mathbf{p}^2}{m^2}$. The first summand in (59) is similar in its form to the current in the Schrödinger equation.

The operators in expressions (58), (59) are diagonal with respect to the upper and lower isotopic components $\left(\Phi_{1,2}\right)_{IFW}$, and, consequently, considering the form of basis functions (30), the current $j^i$ and continuity equation can be written in the chirally symmetric form.

Similarly to (58), (59), one can derive equations of axial current conservation with the matrix $\gamma^5$ between the eight-component spinors $\left(\Phi_{1,2}\right)^{\dagger}_{IFW}$, $\left(\Phi_{1,2}\right)_{IFW}$.

$$\frac{\partial}{\partial t}\left(\left(\Phi_{1,2}\right)^{\dagger}_{IFW}\gamma^5\left(\Phi_{1,2}\right)_{IFW}\right) = -i\left(\left(\Phi_{1,2}\right)^{\dagger}_{IFW}\gamma^5\tau_3 E\left(\Phi_{1,2}\right)_{IFW} - \left(E\left(\Phi_{1,2}\right)^{\dagger}_{IFW}\right)\gamma^5\tau_3\left(\Phi_{1,2}\right)_{IFW}\right) = -div\,\mathbf{j}_A \tag{60}$$

The axial current is preserved in the isotopic Foldy-Wouthuysen representation irrespective of whether fermions possess mass or not.

Recall that in the Dirac representation, the axial current is preserved only for massless fermions.

$$\partial_\mu j^{\mu 5} = 2im\psi^\dagger \beta\gamma^5\psi;\; j^{\mu 5}(x) = \psi^\dagger(x)\alpha^\mu\gamma^5\psi(x). \qquad (61)$$

## 5. Possibility of spontaneous breaking of parity in the isotopic Foldy-Wouthuysen representation

In the previous section, equations (41), (43), (45) and their Hamiltonians were shown to be invariant with respect to chiral symmetry transformations (50).

Hamiltonian density (46) without interaction is actually written separately for left and right particles, and for left and right antiparticles.

Hamiltonian (42) without interaction contains only right fermions and only left antifermions (see Fig. 2).

Hamiltonian (44) without interaction contains only left fermions and only right antifermions (Fig. 3).

Vacuum or the ground state of free Hamiltonian (46) contains a Dirac sea of negative energy states of left and right fermions.

Vacuum of Hamiltonian (42) is a Dirac "sea" of left fermions with $E < 0$, and that of Hamiltonian (44) is a Dirac "sea" of right fermions with $E < 0$.

Since Eqs. (41), (43), (45) were obtained by unitary IFW transformations from equivalent Eqs. (15), (16), (19), it is clear that there exists degenerate vacuum and conditions for spontaneous breaking of P - symmetry.

Probable connection between spontaneous breaking of parity in the IFW representation and the composition of elementary particles of "dark matter" is discussed in Part II of this study.

## 6. Quantum electrodynamics as an example of chirally symmetric theory with massive fermions in the isotopic Foldy-Wouthuysen representation

To construct quantum electrodynamics in the IFW representation, let us use the equation of a sixteen-component field $\Phi_{IFW}(x)$ (45) and Hamiltonian density (46).

### 6.1 Feynman rules in the *IFW* representation

The Feynman propagator of the Dirac equation in the isotopic Foldy-Wouthuysen representation is equal to

$$S^{\Phi}_{IFW}(x-y)=\frac{1}{(2\pi)^4}\int d^4p\frac{e^{-ip(x-y)}}{p_0-E_{16\times16}\tau_3E}=\frac{1}{(2\pi)^4}\int d^4pe^{-ip(x-y)}\frac{p_0+E_{16\times16}\tau_3E}{p^2-m^2+i\varepsilon}=$$
$$=-i\theta(x_0-y_0)\int d\mathbf{p}\sum_s\Phi^{(+)}_{IFW}(x,s)\left(\Phi^{(+)}_{IFW}(y,s)\right)^{\dagger}+i\theta(y_0-x_0)\int d\mathbf{p}\sum_s\Phi^{(-)}_{IFW}(x,s)\left(\Phi^{(-)}_{IFW}(y,s)\right)^{\dagger} \tag{62}$$

Expression (62) implies the Feynman rule of pole bypassing; $\theta(x_0)=\begin{cases}1,\ x_0>0\\0,\ x_0<0\end{cases}$, the functions $\Phi^{(+)}_{IFW}(x,s)$, $\Phi^{(-)}_{IFW}(x,s)$ are defined by formulas (33), (30).

Integral equations for $\Phi_{IFW}(x)$ in accordance with (45), (62) are of the form

$$\Phi_{IFW}(x)=\left(\Phi_0(x)\right)_{IFW}+\int d^4yS^{\Phi}_{IFW}(x-y)\left[\left(K^{\Phi}_1+K^{\Phi}_2+K^{\Phi}_3+\ldots\right)\Phi_{IFW}(y)\right] \tag{63}$$

In expression (63), $\left(\Phi_0(x)\right)_{IFW}$ are solutions of free Dirac equations in the *IFW* representation (see (33), (30)).

Expressions (62), (63) make it possible to set up Feynman rules for writing elements of the scattering matrix $S_{fi}$ and calculating quantum electrodynamics (QED) processes [10]. As distinct from the Dirac representation, in the *IFW* representation there are infinitely many types of photon interaction vertices depending on the order of the perturbation theory: one-photon interaction vertex corresponds to factors $-iK^{\Phi}_{1\mu}$, two-photon interaction vertex corresponds to factors $-iK^{\Phi}_{2\mu\nu}$ and so on. For convenience, the quantities $K^{\Phi}_{1\mu}, K^{\Phi}_{2\mu}\ldots$ denote respective parts of interaction Hamiltonian terms $K^{\Phi}_1, K^{\Phi}_2\ldots$ without electromagnetic potentials $A^{\mu}, A^{\mu}A^{\nu},\ldots\ldots$

Each external fermion line corresponds to one of functions (33). As usual, positive-energy solutions correspond to particles, while negative-energy ones correspond to antiparticles. Other Feynman rules remain the same as in spinor electrodynamics in the Dirac representation.

## 6.2 Calculations of QED processes in the *FW* representation

Some QED processes in the first and second orders of the perturbation theory were considered with the formulated Feynman rules. Calculations were performed for cross sections of Coulomb electron scattering, Moeller scattering, Compton effect, electron-positron pair annihilation; electron self-energy, vacuum polarization, anomalous magnetic momentum of the electron, Lamb shift of atomic energy levels. Below are the Feynman diagrams of the processes under consideration. The calculations are briefly described in Appendix 2.

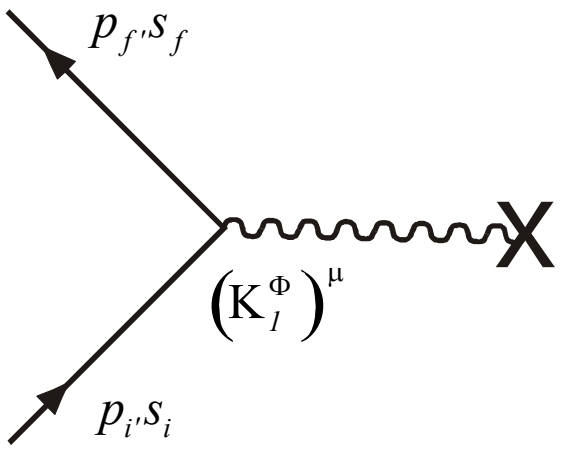

Fig. 4. Electron scattering in the Coulomb field

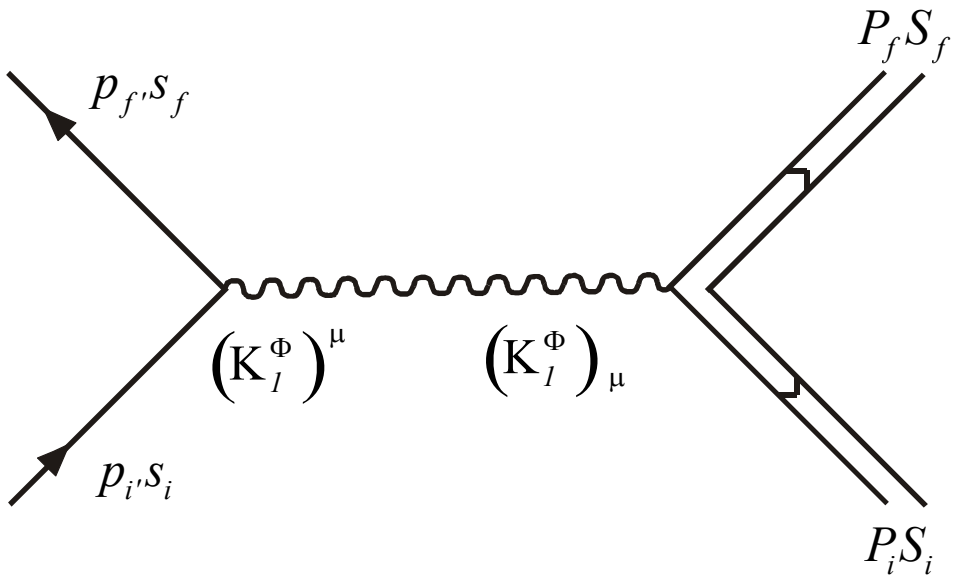

Fig. 5. Electron scattering on on a Dirac proton (Moeller scattering)

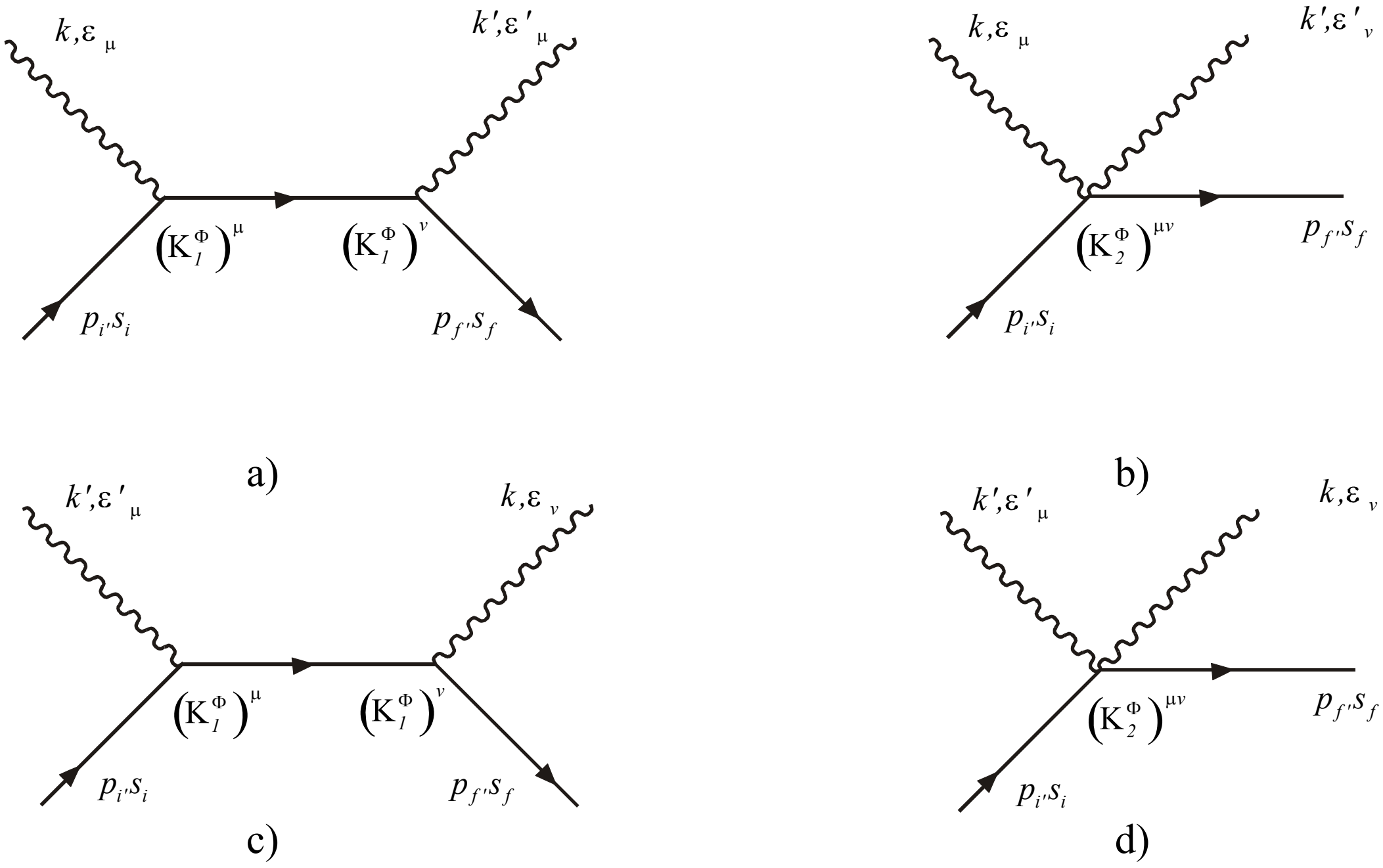

Fig. 6. Compton electron scattering

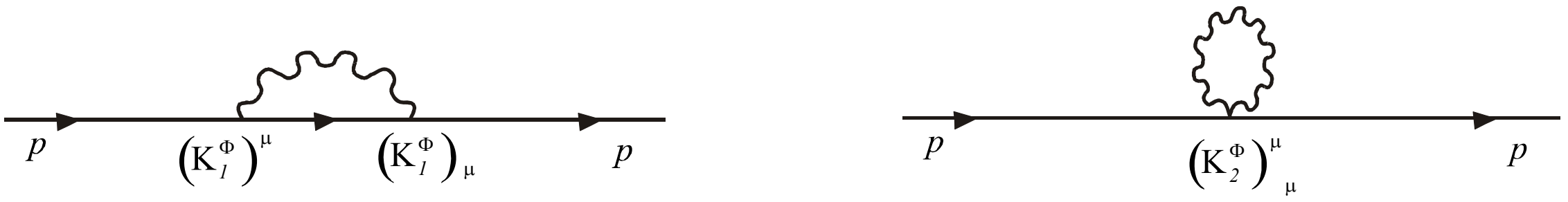

a) b)

Fig. 7. Electron self-energy

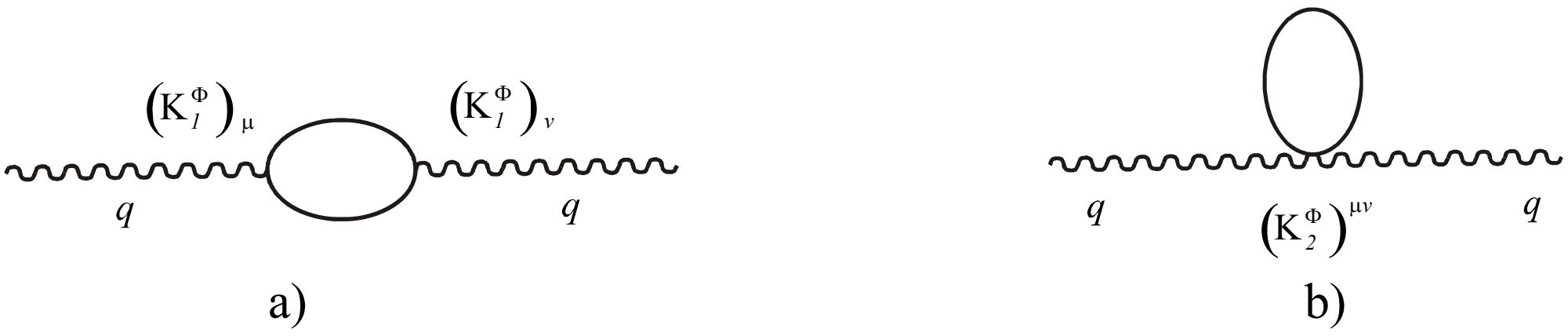

Fig. 8. Vacuum polarization

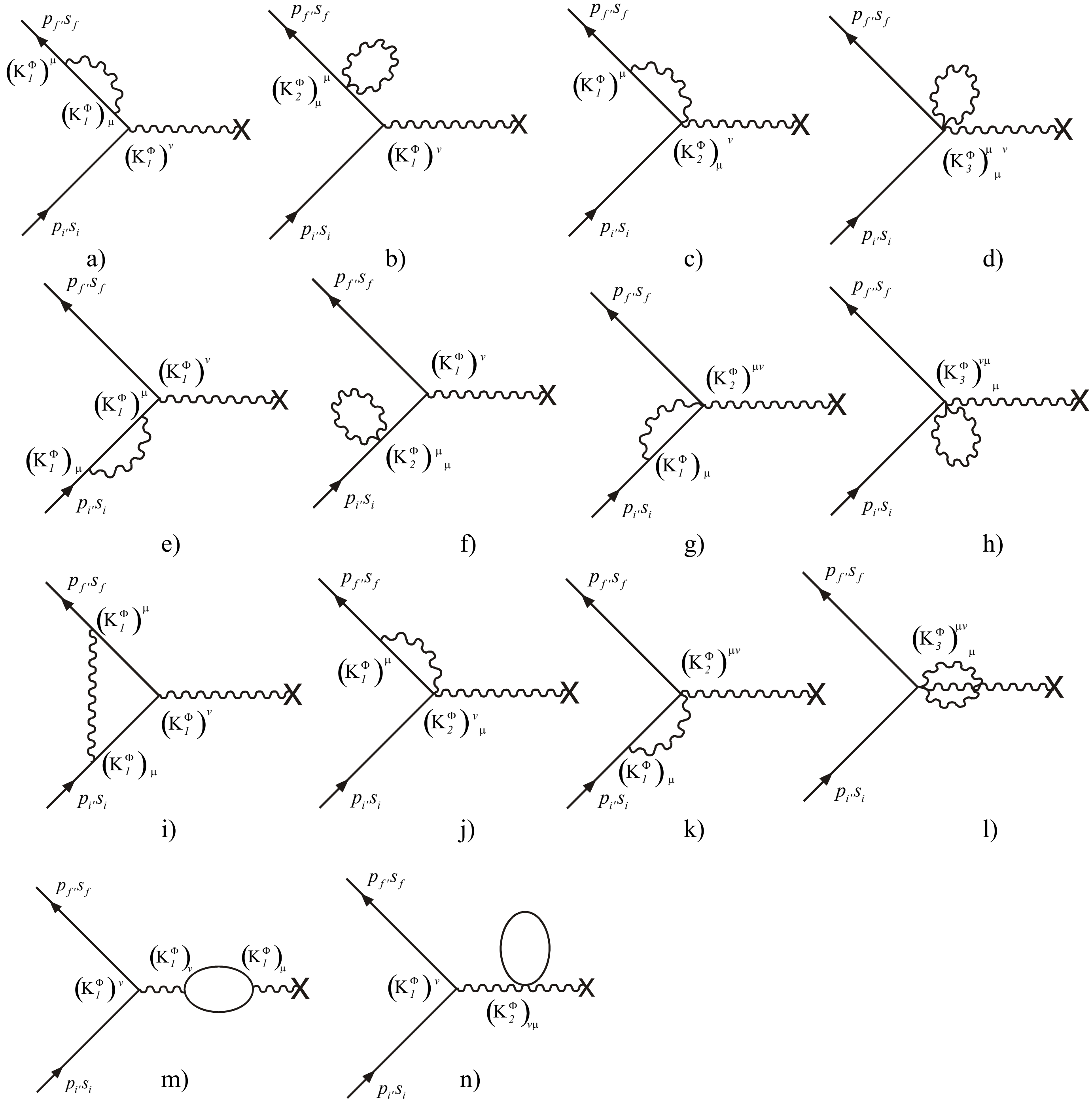

Fig. 9. Radiation corrections to electron scattering in the external field

Final results of the QED process calculations, the diagrams of which are shown in Figs. 4-8, agree with respective quantities calculated in the Dirac representation. Radiation corrections to electron scattering in the external field (Fig. 9) with mass and charge renormalization give a correct value of the anomalous magnetic electron and Lamb shift of electron levels. Appendix 2 also contains results of the electron-positron pair annihilation calculations, which agree with similar results calculated within the standard QED framework.

An important feature of the theory for the case, when external fermion line moments lie on the mass surface $(p^2 = m^2)$, is that the contribution of the diagrams with fermion propagators is compensated by the contribution of respective terms in the diagrams with vertices of higher order expansion in $e$. For example, in Fig. 6, the contribution of diagrams a) and c) is compensated by the corresponding parts of the contribution of diagrams b) and d); the contribution of diagrams a) in Fig. 7 is canceled by the contribution of the corresponding part of diagram b); in Fig. 9, the contribution of diagrams a), b), c) is canceled by the contribution of the corresponding part of diagram d); similar compensation occurs for diagrams e), f), g) and h); i), j), k) and l), respectively. In the case of real external fermions under consideration, vertex operators $K_n^{\Phi}$ are simplified significantly due to the law of energy-moment conservation (see [4], [5] and, e.g., Appendix 1 (A6) – (A12)).

Similar compensation occurs in diagrams a) and b) in Fig. 8 and in diagrams m) and n) in Fig. 9 due to the diagonal matrix summation and equal values of momentum variables at the beginning and at the end of the resulting integrand (see Appendix 2, Sections 5 and 6).

Taking the foregoing into account, one can expand the scattering matrix in the powers of $e$ in such a manner that elements of the matrix $S_{fi}$ will not contain terms with electron-positron propagators. In this case, matrix elements of vertex operators in the momentum representation (see Appendix 1) transform in the following way

$$< \mathbf{p}^{'} \left| K_1^{\Phi} \right| \mathbf{p}^{''} > \to < \mathbf{p}^{'} \left| C^{\Phi} \right| \mathbf{p}^{''} > \tag{64}$$

$$\begin{aligned} < \boldsymbol{p}^{'} \left| K_2^{\Phi} \right| \boldsymbol{p}^{''} > \to \sum_{\nu_1,\nu_2=\pm 1} \int d\boldsymbol{p}^{'''} \frac{1}{R^{'}+R^{''}} \Bigg\{ & \frac{1}{E_{16\times 16}\tau_3 E^{''} - E_{16\times 16}\tau_3 E^{'''} - \nu_2 k_2^0} < \boldsymbol{p}^{'} \left| RC^{\Phi} \right| \boldsymbol{p}^{'''} >< \boldsymbol{p}^{'''} \left| C^{\Phi} \right| \boldsymbol{p}^{''} > + \\ & + \frac{1}{E_{16\times 16}\tau_3 E^{'} - E_{16\times 16}\tau_3 E^{'''} + \nu_1 k_1^0} < \boldsymbol{p}^{'} \left| C^{\Phi} \right| \boldsymbol{p}^{'''} >< \boldsymbol{p}^{'''} \left| C^{\Phi} R \right| \boldsymbol{p}^{''} > + \\ & + \frac{1}{E_{16\times 16}\tau_3 E^{''} + E_{16\times 16}\tau_3 E^{'''} - \nu_2 k_2^0} < \boldsymbol{p}^{'} \left| RN^{\Phi} \right| \boldsymbol{p}^{'''} >< \boldsymbol{p}^{'''} \left| N^{\Phi} \right| \boldsymbol{p}^{''} > + \\ & + \frac{1}{E_{16\times 16}\tau_3 E^{'} + E_{16\times 16}\tau_3 E^{'''} + \nu_1 k_1^0} < \boldsymbol{p}^{'} \left| N^{\Phi} \right| \boldsymbol{p}^{'''} >< \alpha \boldsymbol{p}^{'''} \left| N^{\Phi} R \right| \boldsymbol{p}^{''} > \Bigg\} \end{aligned} \tag{65}$$

$$
\begin{aligned}
&<\boldsymbol{p}^{'}\left|K_3^{\Phi}\right|\boldsymbol{p}^{''}>\to\sum_{\nu_1,\nu_2,\nu_3=\pm 1}\int d\boldsymbol{p}^{'''}\,d\boldsymbol{p}^{IV}\,\frac{1}{R^{'}+R^{''}}\left[\frac{1}{(E_{16\times 16}\tau_3E^{IV}-E_{16\times 16}\tau_3E^{'}-\nu_1k_1^0-\nu_2k_2^0)(E_{16\times 16}\tau_3E^{'''}-E_{16\times 16}\tau_3E^{'}-\nu_1k_1^0)}\times\right.\\
&\times<\boldsymbol{p}^{'}\left|C^{\Phi}\right|\boldsymbol{p}^{'''}><\boldsymbol{p}^{'''}\left|C^{\Phi}\right|\boldsymbol{p}^{IV}><\boldsymbol{p}^{IV}\left|C^{\Phi}R\right|\boldsymbol{p}^{''}>+\\
&+\frac{1}{(E_{16\times 16}\tau_3E^{III}-E_{16\times 16}\tau_3E^{''}+\nu_2k_2^0+\nu_3k_3^0)(E_{16\times 16}\tau_3E^{IV}-E_{16\times 16}\tau_3E^{''}+\nu_3k_3^0)}\times\\
&\times<\boldsymbol{p}^{'}\left|RC^{\Phi}\right|\boldsymbol{p}^{'''}><\boldsymbol{p}^{'''}\left|C^{\Phi}\right|\boldsymbol{p}^{IV}><\boldsymbol{p}^{IV}\left|C^{\Phi}\right|\boldsymbol{p}^{''}>-\\
&-\frac{1}{(E_{16\times 16}\tau_3E^{IV}+E_{16\times 16}\tau_3E^{'}+\nu_1k_1^0+\nu_2k_2^0)(E_{16\times 16}\tau_3E^{'''}-E_{16\times 16}\tau_3E^{'}-\nu_1k_1^0)}\times\\
&\times<\boldsymbol{p}^{'}\left|C^{\Phi}\right|\boldsymbol{p}^{'''}><\boldsymbol{p}^{'''}\left|N^{\Phi}\right|\boldsymbol{p}^{IV}><\boldsymbol{p}^{IV}\left|N^{\Phi}R\right|\boldsymbol{p}^{''}>-\\
&-\frac{1}{(E_{16\times 16}\tau_3E^{''}+E_{16\times 16}\tau_3E^{'''}-\nu_2k_2^0-\nu_3k_3^0)(E_{16\times 16}\tau_3E^{IV}-E_{16\times 16}\tau_3E^{''}+\nu_3k_3^0)}\times\\
&\times<\boldsymbol{p}^{'}\left|RN^{\Phi}\right|\boldsymbol{p}^{'''}><\boldsymbol{p}^{'''}\left|N^{\Phi}\right|\boldsymbol{p}^{IV}><\boldsymbol{p}^{IV}\left|C^{\Phi}\right|\boldsymbol{p}^{''}>+\\
&+\frac{1}{(E_{16\times 16}\tau_3E^{IV}+E_{16\times 16}\tau_3E^{'}+\nu_1k_1^0+\nu_2k_2^0)(E_{16\times 16}\tau_3E^{'''}+E_{16\times 16}\tau_3E^{'}+\nu_1k_1^0)}\times\\
&\times<\boldsymbol{p}^{'}\left|N^{\Phi}\right|\boldsymbol{p}^{'''}><\boldsymbol{p}^{'''}\left|C^{\Phi}\right|\boldsymbol{p}^{IV}><\boldsymbol{p}^{IV}\left|N^{\Phi}R\right|\boldsymbol{p}^{''}>+\\
&+\frac{1}{(E_{16\times 16}\tau_3E^{''}+E_{16\times 16}\tau_3E^{'''}-\nu_2k_2^0-\nu_3k_3^0)(E_{16\times 16}\tau_3E^{IV}+E_{16\times 16}\tau_3E^{''}-\nu_3k_3^0)}\times\\
&\times<\boldsymbol{p}^{'}\left|RN^{\Phi}\right|\boldsymbol{p}^{'''}><\boldsymbol{p}^{'''}\left|C^{\Phi}\right|\boldsymbol{p}^{IV}><\boldsymbol{p}^{IV}\left|N^{\Phi}\right|\boldsymbol{p}^{''}>-\\
&-\frac{1}{(E_{16\times 16}\tau_3E^{IV}-E_{16\times 16}\tau_3E^{'}-\nu_1k_1^0-\nu_2k_2^0)(E_{16\times 16}\tau_3E^{'''}+E_{16\times 16}\tau_3E^{'}+\nu_1k_1^0)}\times\\
&\times<\boldsymbol{p}^{'}\left|N^{\Phi}\right|\boldsymbol{p}^{'''}><\boldsymbol{p}^{'''}\left|N^{\Phi}\right|\boldsymbol{p}^{IV}><\boldsymbol{p}^{IV}\left|C^{\Phi}R\right|\boldsymbol{p}^{''}>-\\
&-\frac{1}{(E_{16\times 16}\tau_3E^{'''}-E_{16\times 16}\tau_3E^{''}+\nu_2k_2^0+\nu_3k_3^0)(E_{16\times 16}\tau_3E^{IV}+E_{16\times 16}\tau_3E^{''}-\nu_3k_3^0)}\times\\
&\left.\times<\boldsymbol{p}^{'}\left|RC^{\Phi}\right|\boldsymbol{p}^{'''}><\boldsymbol{p}^{'''}\left|N^{\Phi}\right|\boldsymbol{p}^{IV}><\boldsymbol{p}^{IV}\left|N^{\Phi}\right|\boldsymbol{p}^{''}>\right]
\end{aligned}
\tag{66}
$$

In Eqs. (64) - (66), the operators $C^{\Phi}$ and $N^{\Phi}$ are defined by formulas (49).

## 7. Conclusions

Because of the presence of the mass term in Dirac equation (9), which is linear in $p^0$, it cannot be written in the chirally symmetric form.

One of the results of this study is that it demonstrates the possibility of writing chirally symmetric equations (separately for left and right fermions) for massive fermion fields. In the Dirac representation, these include integro-differential equations (11) and second-order equations with respect to four-momentum $p^{\mu}$ (13).

In order to obtain $p^0$-linear chirally symmetric equations for massive fermions interacting with vector boson fields, the isotopic representation and the isotopic Foldy-Wouthuysen transformation were defined.

A special isotopic space with isotopic matrices $\tau_i$ was introduced. Dirac equations (15), (16), (19) equivalent in physical consequences for the eight-component fields $\Phi_1, \Phi_2$ and sixteen-component field $\Phi$ were derived within the Dirac representation in the introduced isotopic space. After applying the same isotopic Foldy-Wouthuysen transformation to each equation, desired chirally symmetric equations (41), (43), (45) for massive fermions were obtained.

In the IFW representation, resulting equations and Hamiltonians (41) – (46) are invariant with respect to the chiral transformation $\left(\Phi_{1,2}\right)_{IFW} \rightarrow e^{i\alpha\gamma^5}\left(\Phi_{1,2}\right)_{IFW}$.

This invariance is independent of whether fermions possess mass or not.

In the isotopic Foldy-Wouthuysen representation for massive fermions, both vector and axial currents are preserved. In the Dirac representation, the axial current is preserved only for massless fermions.

Equations (41), (43), (45) differ in their physical content because of the features of the Foldy-Wouthuysen transformation (including the IFW transformation), which restricts the space of possible states of a Dirac particle.

Eq. (45) and Hamiltonian (46) describe both right and left fermions, and right and left antifermions.

Eq. (41) and Hamiltonian (42) describe right fermions and left antifermions, and Eq. (43) with Hamiltonian (44) on the contrary describe left fermions and right antifermions.

The consequences of this is vacuum degeneracy in the isotopic Foldy-Wouthuysen representation and the possibility of spontaneous breaking of P - symmetry.

Because of the unitarity of the isotopic Foldy-Wouthuysen transformation, results of some physical process calculations performed in the lowest orders of the perturbation theory within the framework of chirally symmetric quantum electrodynamics in the IFW representation agree with respective results of standard QED.

## APPENDIX 1

### Definition of Hamiltonian $H^{I}_{IFW}$ in the isotopic Foldy-Wouthuysen representation

With the notation of Refs. [4], [5] for the IFW representation,

$$H^{I}_{IFW} = U_{IFW} H_D U^{\dagger}_{IFW} - iU_{IFW}\frac{\partial}{\partial t}\left(U^{\dagger}_{IFW}\right) = \tau_3 E + K_1 + K_2 + K_3 + \ldots$$

$$U_{IFW} = \left(1 + \delta_1 + \delta_2 + \delta_3 + \ldots\right)\left(U_0\right)_{IFW} \qquad \text{(A1)}$$

In equalities (A1), $K_1, \delta_1 \sim q; K_2, \delta_2 \sim q^2; K_3, \delta_3 \sim q^3$; $q$ is the coupling constant;

$$\left(U_0\right)_{IFW} = R\left(1 + L\right),\ R = \sqrt{\frac{E + \tau_3 \boldsymbol{\alpha}\mathbf{p}}{2E}},\ L = \frac{1}{E + \tau_3 \boldsymbol{\alpha}\mathbf{p}}\tau_3 \tau_1 \beta m.$$

$\left(U_0\right)_{IFW}$ is the IFW transformation operator in the absence of interaction $\left(\left(U_0\right)_{IFW}\left(\boldsymbol{\alpha}\mathbf{p} + \tau_1 \beta m\right)\left(U_0\right)^{\dagger}_{IFW} = \tau_3 E\right)$.

It follows from the unitarity condition $U^{\dagger}_{IFW} U_{IFW} = 1$ that

$$\delta^{\dagger}_1 = -\delta_1;$$

$$\delta^{\dagger}_2 = -\delta_2 + \delta_1 \delta_1; \qquad \text{(A2)}$$

$$\delta^{\dagger}_3 = -\delta_3 + \delta_2 \delta_1 + \delta_1 \delta_2 - \delta_1 \delta_1 \delta_1;$$

..................................................

Taking (A2) into account,

$$\begin{aligned} K_1 &= \delta_1 \tau_3 E - \tau_3 E \delta_1 + i\frac{\partial \delta_1}{\partial t} + C + N \\ K_2 &= \delta_2 \tau_3 E - \tau_3 E \delta_2 + i\frac{\partial \delta_2}{\partial t} - K_1 \delta_1 + \delta_1\left(C + N\right) \\ K_3 &= \delta_3 \tau_3 E - \tau_3 E \delta_3 + i\frac{\partial \delta_3}{\partial t} - K_1 \delta_2 - K_2 \delta_1 + \delta_2\left(C + N\right) \end{aligned} \qquad \text{(A3)}$$

.........................................................................................

In Refs. [4], [5], the following relationship was established between even (denoted by $e$) and odd (denoted by $0$) operators $\delta_i$, which respectively do not link and link upper and lower isotopic components of the wave function

$$\begin{aligned} &\delta^e_1 R + R\delta^e_1 = RL\delta^0_1 - \delta^0_1 LR; \\ &\delta^e_2 R + R\delta^e_2 = RL\delta^0_2 - \delta^0_2 LR - RL\left(\delta_1 \delta_1\right)^0 + R\left(\delta_1 \delta_1\right)^e; \\ &\delta^e_3 R + R\delta^e_3 = RL\delta^0_3 - \delta^0_3 LR - RL\left(\delta_2 \delta_1 + \delta_1 \delta_2 - \delta_1 \delta_1 \delta_1\right)^0 + R\left(\delta_2 \delta_1 + \delta_1 \delta_2 - \delta_1 \delta_1 \delta_1\right)^e; \end{aligned} \qquad \text{(A4)}$$

.........................................................................................

As $K_1, K_2, K_3...$ are even operators by definition, odd operators in (A3) need to be set equal to zero

$$\begin{aligned}
&\delta_1^0\tau_3 E-\beta E\delta_1^0+i\frac{\partial\delta_1^0}{\partial t}+N=0;\\
&\delta_2^0\tau_3 E-\tau_3 E\delta_2^0+i\frac{\partial\delta_2^0}{\partial t}-K_1\delta_1^0+\delta_1^0 C+\delta_1^e N=0;\\
&\delta_3^0\tau_3 E-\tau_3 E\delta_3^0+i\frac{\partial\delta_3^0}{\partial t}-K_1\delta_2^0-K_2\delta_1^0+\delta_2^0 C+\delta_2^e N=0\\
&\ldots\ldots\ldots\ldots\ldots\ldots\ldots\ldots\ldots\ldots\ldots\ldots
\end{aligned}\tag{A5}$$

Following this, the Hamiltonian expansion terms $K_1, K_2, K_3...$ are defined in the following way:

$$\begin{aligned}
&K_1=\delta_1^e\tau_3 E-\tau_3 E\delta_1^e+i\frac{\partial\delta_1^e}{\partial t}+C;\\
&K_2=\delta_2^e\tau_3 E-\tau_3 E\delta_1^e+i\frac{\partial\delta_2^e}{\partial t}-K_1\delta_1^e+\delta_1^e C+\delta_1^0 N;\\
&K_3=\delta_3^e\tau_3 E-\tau_3 E\delta_3^e+i\frac{\partial\delta_3^e}{\partial t}-K_1\delta_2^e-K_2\delta_1^e+\delta_2^e C+\delta_2^0 N;\\
&\ldots\ldots\ldots\ldots\ldots\ldots\ldots\ldots\ldots\ldots\ldots\ldots
\end{aligned}\tag{A6}$$

Operator equalities (A6) together with (A4), (A5) allow expressing the Hamiltonian $H_{IFW}$ as a series in powers of the coupling constant $q$.

In the momentum representation, with expansion of vector boson fields in the Fourier integral with notations of Refs. [4], [5], we have

$$\begin{aligned}
&B_\mu(\mathbf{x},t)=\sum_{v=\pm1}\int B_{\mu k}^{(v)}e^{ivk_0 t}e^{-i\mathbf{kx}}d\mathbf{k},\\
&B_{\mu k}^{(v)}=\begin{cases}B_{\mu k},\ v=1\\ B_{\mu(-k)},\ v=-1\end{cases};
\end{aligned}\tag{A7}$$

$$\langle\mathbf{p}'|e^{i\mathbf{kx}}|\mathbf{p}''\rangle=\delta(\mathbf{p}'-\mathbf{p}''-\mathbf{k})\tag{A8}$$

$$\langle\mathbf{p}'|B_\mu|\mathbf{p}''\rangle=\sum_{v=\pm1}\int B_{\mu(\mathbf{p}''-\mathbf{p}')}^{(v)}e^{ivk_0 t}\tag{A9}$$

$$\langle\mathbf{p}'|B_\mu B_\alpha|\mathbf{p}''\rangle=\sum_{v,v'=\pm1}\int d\mathbf{p}'''B_{\mu(\mathbf{p}'''-\mathbf{p}')}^{(v)}B_{\alpha(\mathbf{p}''-\mathbf{p}''')}^{(v')}e^{i(vk_0+v'k_0)t}\tag{A10}$$

$$\ldots\ldots\ldots\ldots\ldots\ldots\ldots\ldots\ldots\ldots\ldots\ldots$$

Considering (A7) – (A10), we obtain from (A5) that

$$\begin{aligned}
&\langle\mathbf{p}'|\delta_1^0|\mathbf{p}''\rangle=\sum_{v=\pm1}\frac{1}{\tau_3 E'+\tau_3 E''+vk_0}\langle\mathbf{p}'|N|\mathbf{p}''\rangle\\
&\langle\mathbf{p}'|\delta_2^0|\mathbf{p}''\rangle=\sum_{v,v'=\pm1}\frac{1}{\tau_3 E'+\tau_3 E''+vk_0+v'k'_0}\langle\mathbf{p}'|-K_1\delta_1^0+\delta_1^0 C+\delta_1^e N|\mathbf{p}''\rangle\\
&\langle\mathbf{p}'|\delta_3^0|\mathbf{p}''\rangle=\sum_{v,v',v''=\pm1}\frac{1}{\tau_3 E'+\tau_3 E''+vk_0+v'k'_0+v''k''_0}\langle\mathbf{p}'|-K_1\delta_2^0-K_2\delta_1^0+\delta_2^0 C+\delta_2^e N|\mathbf{p}''\rangle
\end{aligned}\tag{П11}$$

Using (A11) and equalities (A4), (A6), one can obtain the sought expansion terms $K_n$ in the momentum representation. For example,

$$\left\langle \mathbf{p}'|K_1|\mathbf{p}''\right\rangle = \sum_{\nu=\pm 1} \frac{\tau_3 E'' - \tau_3 E' - \nu k_0}{R' + R''} \left( -\frac{1}{\tau_3 E' + \tau_3 E' - \nu k_0} \left\langle \mathbf{p}'|RLN|\mathbf{p}''\right\rangle - \right.$$
$$\left. - \frac{1}{\tau_3 E' + \tau_3 E'' + \nu k_0} \left\langle \mathbf{p}'|NLR|\mathbf{p}''\right\rangle \right) + \left\langle \mathbf{p}'|C|\mathbf{p}''\right\rangle \qquad \text{(A12)}$$

## APPENDIX 2

## Calculations of some quantum electrodynamics processes in the isotopic Foldy-Wouthuysen representation

**1. Electron scattering in the Coulomb field** $A_0(x) = -\dfrac{Ze}{4\pi|\boldsymbol{x}|}$

$$S_{fi} = -i\int d^4x \Phi^{(+)\dagger}_{IFW}(x, p_f, s_f)(K_1^{\Phi})^0 \cdot A_0 \Phi^{(+)}_{IFW}(x, p_i, s_i) =$$

$$= -\frac{i\delta(E_f - E_i)}{(2\pi)^2} \Phi^{+\dagger}_{IFW}(p_f, s_f) < \mathbf{p}_f \left|(C^{\Phi})^0 A_0\right| \mathbf{p}_i > \Phi^{(+)}_{IFW}(p_f, s_f) =$$

$$= i\frac{Ze^2}{2\mathbf{q}^2}\frac{\delta(E_f - E_i)}{(2\pi)^2}\left[\Phi^{(+)\dagger}_{1IFW}(p_f, s_f)\sqrt{\frac{E_f + \tau_3\boldsymbol{\alpha}\mathbf{p}_f}{2E_f}}\left(1 + \frac{m^2}{\left(E_f + \tau_3\boldsymbol{\alpha}\mathbf{p}_f\right)\left(E_i + \tau_3\boldsymbol{\alpha}\mathbf{p}_i\right)}\right)\right]\sqrt{\frac{E_i + \tau_3\boldsymbol{\alpha}\mathbf{p}_i}{2E_i}}\Phi^{(+)}_{1IFW}(p_i, s_i) +$$

$$\left. + \Phi^{(+)\dagger}_{2IFW}(p_f, s_f)\sqrt{\frac{E_f + \tau_3\boldsymbol{\alpha}\mathbf{p}_f}{2E_f}}\left(1 + \frac{m^2}{\left(E + \tau_3\boldsymbol{\alpha}\mathbf{p}_f\right)\left(E + \tau_3\boldsymbol{\alpha}\mathbf{p}_i\right)}\right)\sqrt{\frac{E_i + \tau_3\boldsymbol{\alpha}\mathbf{p}_i}{2E_i}}\Phi^{(+)}_{2IFW}(p_i, s_i)\right],$$

$$\mathbf{q} = \mathbf{p}_f - \mathbf{p}_i;\ q^{\Phi} \to \frac{e}{2}E_{16\times 16}$$

The designation $(K_1^{\Phi})^0 \cdot A_0$, made for convenience actually means that $(K_1^{\Phi})^0 \cdot A_0 \equiv K_1^{\Phi}$ with $\mathbf{A}(x) = 0$. That is, $A_0(x)$ put in places defined by the expression for $K_1$ in Appendix 1. The same applies to the designation $(C^{\Phi})^0 \cdot A_0$. The transition from $(K_1^{\Phi})^0 \cdot A_0$ to $(C^{\Phi})^0 \cdot A_0$ is made in accordance with (64).

Then, using the matrix element $S_{fi}$ one can obtain the differential Mott scattering cross-section that takes the form of the Rutherford cross-section in non-relativistic case.

2. **Electron scattering on a Dirac proton (Moeller scattering)**

$$S_{fi} = -i\int d^4x d^4y\, \Phi_{IFW}^{(+)\dagger}(x,p_f,s_f)(K_1^{\Phi})^{\alpha}\Phi_{IFW}^{(+)}(x,p_i,s_i)D_F(x-y)\times$$

$$\times\Phi_{IFW}^{(+)\dagger}(y,P_f,S_f)(-K_1^{\Phi})_{\alpha}\Phi_{IFW}^{(+)}(y,P_i,S_i) =$$

$$= -\frac{i\delta^4\left(P_f - P_i + p_f - p_i\right)}{4(p_f - p_i)^2}2\pi\Big[\Phi_{1IFW}^{(+)\dagger}(p_f,s_f)\left\langle \boldsymbol{p}_f\left|(C^{\Phi})^{\alpha}\right|\boldsymbol{p}_i\right\rangle\Phi_{1IFW}^{(+)}(p_i,s_i)+$$

$$+\Phi_{2IFW}^{(+)\dagger}(p_f,s_f) < \boldsymbol{p}_f\left|(C^{\Phi})^{\alpha}\right|\boldsymbol{p}_i > \Phi_{2IFW}^{(+)}(p_i,s_i)\Big]\times$$

$$\times\Big[\Phi_{1IFW}^{(+)\dagger}(P_f,S_f) < \boldsymbol{P}_f\left|C_{\alpha}^{\Phi}\right|\boldsymbol{P}_i > \Phi_{1IFW}^{(+)}(P_i,S_i)+$$

$$+\Phi_{2IFW}^{(+)\dagger}(P_f,S_f) < \boldsymbol{P}_f\left|C_{\alpha}^{\Phi}\right|\boldsymbol{P}_i > \Phi_{2IFW}^{(+)}(P_i,S_i)\Big]$$

$D_F(x-y)$ is a standard photon propagator; $q^{\Phi} \to \frac{e}{2}E_{16\times16}$.

The matrix element $S_{f_i}$ makes it possible to determine the Moeller electron scattering cross-section.

3. **Compton electron scattering**

$$S_{fi} = -i\Phi_{IFW}^{(+)\dagger}(p_f,s_f)\Big\{\int\frac{d^4z d^4y d^4p_1}{(2\pi)^7\sqrt{2k_0^{'}2k_0}}(e^{ip_f y}K_{1\mu}^{\Phi}\varepsilon'^{\mu}e^{ik'y}\frac{e^{-ip_1 y}}{p_1^0 - E_{16\times16}\tau_3 E(\mathbf{p}_1)}\cdot e^{ip_1 z}\times$$

$$\times K_{1\nu}^{\Phi}\varepsilon^{\nu}e^{-ikz}e^{-ip_i z} + e^{ip_f y}K_{1\mu}^{\Phi}\varepsilon^{\mu}e^{-iky}\frac{e^{-ip_1 y}}{p_1^0 - E_{16\times16}\tau_3 E(\mathbf{p}_1)}e^{ip_1 z}K_{1\nu}^{\Phi}\varepsilon'^{\nu}e^{ik'z}e^{-ip_i z})+$$

$$+\int d^4y\frac{1}{(2\pi)^3\sqrt{2k_0^{'}\cdot 2k_0}}(e^{ip_f y}K_{2\mu\nu}^{\Phi}\varepsilon'^{\mu}e^{ik'y}\varepsilon^{\nu}e^{-iky}e^{-ip_i y}+$$

$$+e^{ip_f y}K_{2\mu\nu}^{\Phi}\varepsilon^{\mu}e^{-iky}\varepsilon'^{\nu}e^{-ip_i y})\Big\}\Phi_{IFW}^{(+)}(p_i,s_i)$$

The first integral combines the contributions of diagrams a) and c) in Fig. 6; and the second one combines the contributions of diagrams b) and d) in Fig.6.

The designation $K_{1\mu}^{\Phi}\varepsilon^{\mu}e^{-iky}$, $K_{2\mu\nu}^{\Phi}\varepsilon^{\mu}e^{-iky}\varepsilon'^{\nu}e^{ik'y}$ and so on is understood to have the same meaning as in Section 1 above.

As noted in Subsection 6.2, the contribution to the matrix element $S_{fi}$ is made only by the summands in expression (65) for $K_2^{\Phi}$. Taking the foregoing into account,

$$S_{fi}=\frac{-i(2\pi)^4\cdot\delta^4(p_i+k-p_f-k')}{(2\pi)^3\sqrt{2k_0\cdot 2k_0^{'}}}\Big[\Phi_{1IFW}^{(+)\dagger}(p_f,s_f)A\Phi_{IFW}^{(+)}(p_i,s_i)+$$
$$+\Phi_{2IFW}^{(+)\dagger}(p_f,s_f)A\Phi_{2IFW}^{(+)}(p_i,s_i)\Big],$$

where

$$A=\Bigg[C_\mu(\mathbf{p}_f;\mathbf{p}_i+\mathbf{k})\varepsilon^{'\mu}\frac{\frac{1}{2}(1+\tau_3)}{\tau_3E(\mathbf{p}_i)+k_0-\tau_3E(\mathbf{p}_i+\mathbf{k})}C_\mu(\mathbf{p}_i+\mathbf{k};\mathbf{p}_i)\varepsilon^\mu+$$
$$+N_\mu(\mathbf{p}_f;\mathbf{p}_i+\mathbf{k})\varepsilon^{'\mu}\frac{\frac{1}{2}(1-\tau_3)}{-\tau_3E(\mathbf{p}_i)+k_0+\tau_3E(\mathbf{p}_i+\mathbf{k})}N_\mu(\mathbf{p}_i+\mathbf{k};\mathbf{p}_i)\varepsilon^\mu+$$
$$+C_\mu(\mathbf{p}_f;\mathbf{p}_i-\mathbf{k}')\varepsilon^\mu\frac{\frac{1}{2}(1+\tau_3)}{\tau_3\cdot E(\mathbf{p}_i)-k_0'-\tau_3E(\mathbf{p}_i-\mathbf{k}')}C_\mu(\mathbf{p}_i-\mathbf{k}';\mathbf{p}_i)\varepsilon'^\mu+$$
$$+N_\mu(\mathbf{p}_f;\mathbf{p}_i-\mathbf{k}')\varepsilon^\mu\frac{\frac{1}{2}(1-\tau_3)}{-\tau_3E(\mathbf{p}_i)-k_0'+\tau_3E(\mathbf{p}_i-\mathbf{k}')}N_\mu(\mathbf{p}_i-\mathbf{k}';\mathbf{p}_i)\varepsilon'^\mu\Bigg]+$$

$$+\Bigg[C_\mu(\mathbf{p}_f;\tau_1;\mathbf{p}_i+\mathbf{k})\varepsilon^{'\mu}\frac{\frac{1}{2}(1+\tau_3)}{\tau_3E(\mathbf{p}_i)+k_0-\tau_3E(\mathbf{p}_i+\mathbf{k})}C_\mu(\mathbf{p}_i+\mathbf{k};\tau_1;\mathbf{p}_i)\varepsilon^\mu+$$
$$+N_\mu(\mathbf{p}_f;\tau_1;\mathbf{p}_i+\mathbf{k})\varepsilon^{'\mu}\frac{\frac{1}{2}(1-\tau_3)}{-\tau_3E(\mathbf{p}_i)+k_0+\tau_3E(\mathbf{p}_i+\mathbf{k})}N_\mu(\mathbf{p}_i+\mathbf{k};\tau_1;\mathbf{p}_i)\varepsilon^\mu+$$
$$+C_\mu(\mathbf{p}_f;\tau_1;\mathbf{p}_i-\mathbf{k}')\varepsilon^\mu\frac{\frac{1}{2}(1+\tau_3)}{\tau_3E(\mathbf{p}_i)-k_0^{'}-\tau_3E(\mathbf{p}_i-\mathbf{k}')}C_\mu(\mathbf{p}_i-\mathbf{k}';\tau_1;\mathbf{p}_i)\varepsilon^{'\mu}+$$
$$+N_\mu(\mathbf{p}_f;\tau_1;\mathbf{p}_i-\mathbf{k}')\varepsilon^\mu\frac{\frac{1}{2}(1-\tau_3)}{-\tau_3E(\mathbf{p}_i)-k_0^{'}+\tau_3E(\mathbf{p}_i-\mathbf{k}')}N_\mu(\mathbf{p}_i-\mathbf{k}';\tau_1;\mathbf{p}_i)\varepsilon^{'\mu}\Bigg];$$

$$C^\mu(\mathbf{p}_f;\mathbf{p}_i+\mathbf{k})=\begin{cases}\frac{1}{2}eR_f\left(1-\frac{1}{E(\mathbf{p}_f)+\tau_3\boldsymbol{\alpha}\mathbf{p}_f}\tau_3\tau_1\beta m\frac{1}{E(\mathbf{p}_i+\mathbf{k})+\tau_3\boldsymbol{\alpha}(\mathbf{p}_i+\mathbf{k})}\tau_3\tau_1\beta m\right)R_1; & \mu=0\\ -\frac{1}{2}eR_f\left(\alpha^k-\frac{1}{E(\mathbf{p}_f)+\tau_3\boldsymbol{\alpha}\mathbf{p}_f}\tau_3\tau_1\beta m\alpha^k\frac{1}{E(\mathbf{p}_i+\mathbf{k})+\tau_3\boldsymbol{\alpha}(\mathbf{p}_i+\mathbf{k})}\tau_3\tau_1\beta m\right)R_1; & \mu=k;\quad k=1,2,3\end{cases}$$

$$N^\mu(\mathbf{p}_f;\mathbf{p}_i+\mathbf{k})=\begin{cases}\frac{1}{2}eR_f\left(\frac{1}{E(\mathbf{p}_f)+\tau_3\boldsymbol{\alpha}\mathbf{p}_f}\tau_3\tau_1\beta m-\frac{1}{E(\mathbf{p}_i+\mathbf{k})+\tau_3\boldsymbol{\alpha}(\mathbf{p}_i+\mathbf{k})}\tau_3\tau_1\beta m\right)R_1; & \mu=0\\ -\frac{1}{2}eR_f\left(\frac{1}{E(\mathbf{p}_f)+\tau_3\boldsymbol{\alpha}\mathbf{p}_f}\tau_3\tau_1\beta m\alpha^k-\alpha^k\frac{1}{E(\mathbf{p}_i+\mathbf{k})+\tau_3\boldsymbol{\alpha}(\mathbf{p}_i+\mathbf{k})}\tau_3\tau_1\beta m\right)R_1; & \mu=k;\quad k=1,2,3\end{cases}$$

$$C^{\mu}(\mathbf{p}_f;\tau_1;\mathbf{p}_i+\mathbf{k})=\begin{cases}\frac{1}{2}eR_f\left(\tau_1-\frac{1}{E(\mathbf{p}_f)+\tau_3\boldsymbol{\alpha}\mathbf{p}_f}\tau_3\tau_1\beta m\tau_1\frac{1}{E(\mathbf{p}_i+\mathbf{k})+\tau_3\boldsymbol{\alpha}(\mathbf{p}_i+\mathbf{k})}\tau_3\tau_1\beta m\right)R_1; \quad \mu=0\\ -\frac{1}{2}eR_f\left(\tau_1\alpha^k-\frac{1}{E(\mathbf{p}_f)+\tau_3\boldsymbol{\alpha}\mathbf{p}_f}\tau_3\tau_1\beta m\tau_1\alpha^k\frac{1}{E(\mathbf{p}_i+\mathbf{k})+\tau_3\boldsymbol{\alpha}(\mathbf{p}_i+\mathbf{k})}\tau_3\tau_1\beta m\right)R_1; \quad \mu=k;\ k=1,2,3\end{cases}$$

$$N^{\mu}(\mathbf{p}_f;\tau_1;\mathbf{p}_i+\mathbf{k})=\begin{cases}\frac{1}{2}eR_f\left(\frac{1}{E(\mathbf{p}_f)+\tau_3\boldsymbol{\alpha}\mathbf{p}_f}\tau_3\tau_1\beta m\tau_1-\tau_1\frac{1}{E(\mathbf{p}_i+\mathbf{k})+\tau_3\boldsymbol{\alpha}(\mathbf{p}_i+\mathbf{k})}\tau_3\tau_1\beta m\right)R_1; \quad \mu=0\\ -\frac{1}{2}eR_f\left(\frac{1}{E(\mathbf{p}_f)+\tau_3\boldsymbol{\alpha}\mathbf{p}_f}\tau_3\tau_1\beta m\alpha^k\tau_1-\tau_1\alpha^k\frac{1}{E(\mathbf{p}_i+\mathbf{k})+\tau_3\boldsymbol{\alpha}(\mathbf{p}_i+\mathbf{k})}\tau_3\tau_1\beta m\right)R_1; \quad \mu=k;\ k=1,2,3\end{cases}$$

$$R_f\equiv R_{p_f}=\sqrt{\frac{E(\mathbf{p}_f)+\tau_3\boldsymbol{\alpha}\mathbf{p}_f}{2E(\mathbf{p}_f)}},\ \ R_1\equiv R_{p_i+k}=\sqrt{\frac{E(\mathbf{p}_i+\mathbf{k})+\tau_3\boldsymbol{\alpha}(\mathbf{p}_i+\mathbf{k})}{2E(\mathbf{p}_i+\mathbf{k})}} \quad \text{и так далее.}$$

Calculations show that the sum of the summands in the first square brackets in the expression for $A$ coincides with the similar sum in the second square brackets.

If we choose a special gauge with transversely polarized initial and final photons in the laboratory frame of reference $(\mathbf{p}_i=0;\quad \varepsilon^0=\varepsilon'^0=0;\quad \boldsymbol{\varepsilon}k=\boldsymbol{\varepsilon}'\mathbf{k}'=0)$, the expression for $A$ will be simplified:

$$A=\frac{e^2}{2}\sqrt{\frac{E(\mathbf{p}_f)+\tau_3\boldsymbol{\alpha}\mathbf{p}_f}{4E(\mathbf{p}_f)}}\left[-\frac{E(\mathbf{p}_f)+\tau_3\boldsymbol{\alpha}\mathbf{p}_f-m}{2m^2}\left(\frac{\boldsymbol{\alpha\varepsilon}'\,\boldsymbol{\alpha}\mathbf{k}\,\boldsymbol{\alpha\varepsilon}}{k_0}+\frac{\boldsymbol{\alpha\varepsilon}\,\boldsymbol{\alpha}\mathbf{k}'\,\boldsymbol{\alpha\varepsilon}'}{k_0'}\right)+\right.$$
$$\left.+\frac{E(\mathbf{p}_f)-\tau_3\boldsymbol{\alpha}\mathbf{p}_f+m}{m^2}\boldsymbol{\varepsilon\varepsilon}'\right].$$

Then, using conventional procedures, one can derive the Klein-Nishina-Tamm formula for the differential Compton scattering cross-section.

### 4. Electron self-energy

$$-i\sum^{(2)}(p)=-\int\frac{d^4k}{(2\pi)^4\cdot k^2}\Big[C^{\mu}(\mathbf{p};\mathbf{p}-\mathbf{k})\frac{\frac{1}{2}(1+\tau_3)}{\tau_3E(\mathbf{p})-k_0-\tau_3E(\mathbf{p}-\mathbf{k})}C_{\mu}(\mathbf{p}-\mathbf{k};\mathbf{p})+$$
$$+C^{\mu}(\mathbf{p};\tau_1;\mathbf{p}-\mathbf{k})\frac{\frac{1}{2}(1+\tau_3)}{\tau_3E(\mathbf{p})-k_0-\tau_3E(\mathbf{p}-\mathbf{k})}C_{\mu}(\mathbf{p}-\mathbf{k};\tau_1;\mathbf{p})+$$
$$+N^{\mu}(\mathbf{p};\mathbf{p}-\mathbf{k})\frac{\frac{1}{2}(1-\tau_3)}{-\tau_3E(\mathbf{p})-k_0+\tau_3E(\mathbf{p}-\mathbf{k})}N_{\mu}(\mathbf{p}-\mathbf{k};\mathbf{p})+$$
$$+N^{\mu}(\mathbf{p};\tau_1;\mathbf{p}-\mathbf{k})\frac{\frac{1}{2}(1-\tau_3)}{-\tau_3E(\mathbf{p})-k_0+\tau_3E(\mathbf{p}-\mathbf{k})}N_{\mu}(\mathbf{p}-\mathbf{k};\tau_1;\mathbf{p})\Big]$$

For the case $p^2=m^2$, considering that the operator $-i\sum^2(p)$ is bracketed between the wave functions $\Phi_{IFW}^{(+)\dagger}$ and $\Phi_{IFW}^{(+)}$, i.e. in accordance with (33) and (45) it is actually bracketed between the functions for right $\left(\Phi^{(+)\dagger}_{1IFW}\ and\ \Phi^{(+)}_{1IFW}\right)$ and left $\left(\Phi^{(+)\dagger}_{2IFW}\ and\ \Phi^{(+)}_{2IFW}\right)$ electrons, we have $-i\sum^{(2)}(p) = -\frac{2e^2}{E(\mathbf{p})}\int\frac{d^4k}{(2\pi)^4\cdot k^2}\cdot\frac{p\cdot k+m^2}{[(p-k)^2-m^2]}$, which agrees with the expression for the mass operator in the Dirac representation taking into account spinor normalization in external electron lines.

### 5. Vacuum polarization

As a result, the following expression for the polarized operator corresponds to the diagrams in Fig. 8:

$$\begin{aligned}\Pi^{\mu\nu}(q) = -\frac{1}{2}\int\frac{d\mathbf{p}}{(2\pi)^3}\Bigg\{&Sp\frac{1}{E(\mathbf{p})-q^0+E(\mathbf{p}-\mathbf{q})}N^{\mu}(\mathbf{p};\mathbf{p}-\mathbf{q})\frac{1-\tau_3}{2}N^{\nu}(\mathbf{p}-\mathbf{q};\mathbf{p})+\\ &+Sp\frac{1}{E(\mathbf{p})+q^0+E(\mathbf{p}-\mathbf{q})}N^{\mu}(\mathbf{p}-\mathbf{q};\mathbf{p})\frac{1-\tau_3}{2}N^{\nu}(\mathbf{p};\mathbf{p}-\mathbf{q})+\\ &+Sp\frac{1}{E(\mathbf{p})-q^0+E(\mathbf{p}-\mathbf{q})}C^{\mu}(\mathbf{p};\tau_1;\mathbf{p}-\mathbf{q})\frac{1-\tau_3}{2}C^{\nu}(\mathbf{p}-\mathbf{q};\tau_1;\mathbf{p})+\\ &+Sp\frac{1}{E(\mathbf{p})+q^0+E(\mathbf{p}-\mathbf{q})}C^{\mu}(\mathbf{p}-\mathbf{q};\tau_1;\mathbf{p})\frac{1-\tau_3}{2}C^{\nu}(\mathbf{p};\tau_1;\mathbf{p}-\mathbf{q})\Bigg\}\end{aligned}$$

After calculating of matrix traces, the expression for $\Pi^{\mu\nu}$ coincides with the Heitler induction tensor $-L^{\mu\nu}$ [11], which in turn coincides with the polarized operator of standard QED.

### 6. Radiation corrections to electron scattering in the external field

When calculating the radiation corrections based on the diagrams of Fig.9, similarly to the cases above, the matrix elements $S_{fi}$ corresponding to the diagrams with electron-positron propagators are cancelled by the parts of the matrix elements corresponding to the propagator-free diagrams in Fig.9 d), h), l). As a result, taking into account the Heitler limiting process for singular denominators [11], the matrix element for the sought radiation corrections takes the following form:

$$S_{fi} = S_{1fi} + S_{2fi};$$

$$S_{1fi} = S^{(1)}_{1fi} + S^{(2)}_{1fi} + S^{(3)}_{1fi} + S^{(4)}_{1fi} + S^{(5)}_{1fi};$$

$$S_{2fi} = S^{(1)}_{2fi} + S^{(2)}_{2fi} + S^{(3)}_{2fi} + S^{(4)}_{2fi} + S^{(5)}_{2fi};$$

$$S^{(1)}_{1fi} = \frac{1}{(2\pi)^3}\Phi^{(+)\dagger}_{1IFW}(p_f,s_f)\left\{\int\frac{d^4k}{(2\pi)^4\cdot k^2}\int\frac{d\varepsilon\delta(\varepsilon)}{\mathbf{p}_f^2\cdot\varepsilon(2+\varepsilon)}\Big[C_\mu(\mathbf{p}_f;\mathbf{p}_f(1+\varepsilon)-\mathbf{k})\times\right.$$

$$\times\frac{\tau_3E(\mathbf{p}_f)+\tau_3E(\mathbf{p}_f(1+\varepsilon))}{\tau_3E(\mathbf{p}_f)-k^0-\tau_3E(\mathbf{p}_f(1+\varepsilon)-\mathbf{k})}C^\mu(\mathbf{p}_f(1+\varepsilon)-\mathbf{k};\mathbf{p}_f(1+\varepsilon))+N_\mu(\mathbf{p}_f;\mathbf{p}_f(1+\varepsilon)-\mathbf{k})\times$$

$$\times\frac{\tau_3E(\mathbf{p}_f)+\tau_3E(\mathbf{p}_f(1+\varepsilon))}{\tau_3E(\mathbf{p}_f)-k^0-\tau_3E(\mathbf{p}_f(1+\varepsilon)-\mathbf{k})}N^\mu(\mathbf{p}_f(1+\varepsilon)-\mathbf{k};\mathbf{p}_f(1+\varepsilon))\Big]C^\nu(\mathbf{p}_f;\mathbf{p}_i)A_\nu(q)+$$

$$+\int\frac{d^4k}{(2\pi)^4\cdot k^2}\int\frac{d\varepsilon\delta(\varepsilon)}{\mathbf{p}_i^2\cdot\varepsilon(2+\varepsilon)}\Big[C_\mu(\mathbf{p}_i(1+\varepsilon);\mathbf{p}_i(1+\varepsilon)-\mathbf{k})\times$$

$$\times\frac{\tau_3E(\mathbf{p}_i)+\tau_3E(\mathbf{p}_i(1+\varepsilon))}{\tau_3E(\mathbf{p}_i)-k^0-\tau_3E(\mathbf{p}_i(1+\varepsilon)-\mathbf{k})}C^\mu(\mathbf{p}_i(1+\varepsilon)-\mathbf{k};\mathbf{p}_i)+N_\mu(\mathbf{p}_i(1+\varepsilon);\ \mathbf{p}_i(1+\varepsilon)-\mathbf{k})\times$$

$$\times\frac{\tau_3E(\mathbf{p}_i)+\tau_3E(\mathbf{p}_i(1+\varepsilon))}{\tau_3E(\mathbf{p}_i)-k^0-\tau_3E(\mathbf{p}_i(1+\varepsilon)-\mathbf{k})}N^\mu(\mathbf{p}_i(1+\varepsilon)-\mathbf{k};\mathbf{p}_i)\Big]C^\nu(\mathbf{p}_f;p_i)A_\nu(q)-$$

$$-\int\frac{d^4k}{(2\pi)^4\cdot k^2}\frac{1}{2\tau_3E(\mathbf{p}_f)}\Big[\frac{1}{\tau_3E(\mathbf{p}_f)-k^0+\tau_3E(\mathbf{p}_f-\mathbf{k})}N_\mu(\mathbf{p}_f;\mathbf{p}_f-\mathbf{k})\,C^\mu(\mathbf{p}_f-\mathbf{k};\mathbf{p}_f)+$$

$$+\frac{1}{\tau_3E(\mathbf{p}_f)-k^0-\tau_3E(\mathbf{p}_f-\mathbf{k})}C_\mu(\mathbf{p}_f;\mathbf{p}_f-\mathbf{k})N^\mu(\mathbf{p}_f-\mathbf{k};\mathbf{p}_f)\Big]N^\nu(\mathbf{p}_f;\mathbf{p}_i)A_\nu(q)-$$

$$-\int\frac{d^4k}{(2\pi)^4\cdot k^2}\frac{1}{2\tau_3E(\mathbf{p}_i)}N^\nu(\mathbf{p}_f;\mathbf{p}_i)A_\nu(q)\Big[\frac{1}{\tau_3E(\mathbf{p}_i)-k^0-\tau_3E(\mathbf{p}_i-\mathbf{k})}\times$$

$$\times N_\mu(\mathbf{p}_i;\mathbf{p}_i-\mathbf{k})\,C^\mu(\mathbf{p}_i-\mathbf{k}_i;\mathbf{p}_i)+\frac{1}{\tau_3E(\mathbf{p}_i)-k^0+\tau_3E(\mathbf{p}_i-\mathbf{k})}\times$$

$$\times C_\mu(\mathbf{p}_i;\mathbf{p}_i-\mathbf{k})N^\mu(\mathbf{p}_i-\mathbf{k}_i;\mathbf{p}_i)\Big]-$$

$$-\int\frac{d^4k}{(2\pi)^4\cdot k^2}\Big[C^\mu(\mathbf{p}_f;\mathbf{p}_f-\mathbf{k})\frac{1}{\tau_3E(\mathbf{p}_f)-k^0-\tau_3E(\mathbf{p}_f-\mathbf{k})}C^\nu(\mathbf{p}_f-\mathbf{k};\mathbf{p}_i-\mathbf{k})A_\nu(q)\times$$

$$\times\frac{1}{\tau_3E(\mathbf{p}_i)-k^0-\tau_3E(\mathbf{p}_i-\mathbf{k})}C_\mu(\mathbf{p}_i-\mathbf{k};\mathbf{p}_i)+$$

$$+N^\nu(\mathbf{p}_f;\mathbf{p}_f-\mathbf{k})\frac{1}{\tau_3E(\mathbf{p}_f)-k^0+\tau_3E(p_f-\mathbf{k})}C^\nu(\mathbf{p}_f-\mathbf{k};\mathbf{p}_i-\mathbf{k})A_\nu(q)\times$$

$$\times\frac{1}{\tau_3E(\mathbf{p}_i)-k^0+\tau_3E(\mathbf{p}_i-\mathbf{k})}N_\mu(\mathbf{p}_i-\mathbf{k};\mathbf{p}_i)+$$

$$+N^\mu(\mathbf{p}_f;\mathbf{p}_f-\mathbf{k})\frac{1}{\tau_3E(\mathbf{p}_f)-k^0+\tau_3E(\mathbf{p}_f-\mathbf{k})}N^\nu(\mathbf{p}_f-\mathbf{k};\mathbf{p}_i-\mathbf{k})A_\nu(q)\times$$

$$\times\frac{1}{\tau_3E(\mathbf{p}_i)-k^0-\tau_3E(\mathbf{p}_i-\mathbf{k})}C_\mu(\mathbf{p}_i-\mathbf{k};\mathbf{p}_i)+$$

$$+C^{\mu}(\mathbf{p}_f;\mathbf{p}_i-\mathbf{k})\frac{1}{\tau_3 E(\mathbf{p}_f)-k^0-\tau_3 E(\mathbf{p}_f-\mathbf{k})}N^{\nu}(\mathbf{p}_f-\mathbf{k};\mathbf{p}_i-\mathbf{k})A_{\nu}(q)\times$$

$$\times\frac{1}{\tau_3 E(\mathbf{p}_i)-k^0+\tau_3 E(\mathbf{p}_i-\mathbf{k})}N_{\mu}(\mathbf{p}_i-\mathbf{k};\mathbf{p}_i)\Bigg]\Bigg\}\Phi^{(+)}_{1IFW}(p_i,s_i).$$

As the calculations show, the expressions for $S^{(2)}_{1fi}$, $S^{(3)}_{1fi}$, $S^{(4)}_{1fi}$ are equal to $S^{(1)}_{1fi}$ and their form differs in the presence in each summand $S^{(1)}_{1fi}$ of two operators ...C...C.., or ...C...N.., or ...N...C.., or ...N...N.. with matrix $\tau_1$ in each of them.

The expression $S^{(5)}_{1fi}$ equals

$$S^{(5)}_{1fi}=-\frac{1}{(2\pi)^3}\Phi^{(+)}_{1IFW}(p_f,s_f)C_{\mu}(\boldsymbol{p}_f,\boldsymbol{p}_i)\frac{\Pi^{\mu\nu}(q)}{q^2}A_{\nu}(q)\Phi^{(+)}_{1IFW}(p_i,s_i).$$

The expression for $S_{2fi}$ equals $S_{1fi}$ and differs formally in the wave functions: Instead of the functions $\Phi^{(+)}_{1IFW}$ for right electrons, it uses the functions $\Phi^{(+)}_{2IFW}$ for left electrons.

After renormalization of electron mass and charge, the resulting matrix element $S_{fi}$ makes it possible to calculate the anomalous magnetic momentum on the electron and the Lamb shift of atomic levels. Final results of the calculations agree with the same quantities in the Dirac representation.

## 7. Electron-positron pair annihilation

The process of electron-positron pair annihilation in the second order of the perturbation theory is represented by the diagram in Fig.6 for Compton electron scattering with the replacement of $\varepsilon,k\to\varepsilon_1,-k_1;\ \varepsilon',k'\to\varepsilon_2,k_2;\ p_i,s_i\to p_-,s_-;\ p_f,s_f\to -p_+,s_+$.

Considering this replacement, the matrix element $S_{\pm}$ of the process is equal to

$$S_{\pm}=\frac{-i(2\pi)^4\delta^4(p_-+p_+-k_1-k_2)}{(2\pi)^3\sqrt{2k_1^0 2k_2^0}}\Big(\Phi^{(-)\dagger}_{2IFW}(-p_+,s_+)A_1\Phi^{(-)}_{1IFW}(p_-,s_-)\times$$
$$\times\Phi^{(-)\dagger}_{1IFW}(-p_+,s_+)A_1\Phi^{(-)}_{2IFW}(p_-,s_-)\Big)$$

Subject to the above replacement, the operator $A_1$ in its structure coincides with the operator $A$ in the expression for $S_{fi}$ for Compton electron scattering, except the position of matrices $\tau_1$ in the operators $C^{\mu}$, $N^{\mu}$.

In the expression for $S_{fi}$, the first four summands contain no matrix $\tau_1$, and the remaining four summands contain $\tau_1$ in each operator $C^{\mu}$, $N^{\mu}$.

In our case, in the expression for $S_{\pm}$, in order to link the right electron and right positron, or the left electron and left positron, each of the eight summands contains only one matrix $\tau_{l}$ located in the operators $C^{\mu}$, $N^{\mu}$ in all possible combinations.

The resulting expression for $S_{\pm}$ allows us to obtain the differential electron-positron pair annihilation cross-section, which coincides with the same cross-section calculated in the Dirac representation.